\documentclass[useAMS,usenatbib]{mn2e}

\usepackage{epsfig}
\usepackage{subfigure}

\newcommand{\be}{\begin{equation}}
\newcommand{\ee}{\end{equation}}
\newcommand{\bea}{\begin{eqnarray}}
\newcommand{\eea}{\end{eqnarray}}

\title[Hubble Diagram Dispersion From Large-Scale
  Structure]{Hubble Diagram Dispersion From Large-Scale Structure}
\author[Timothy Clifton and Joe Zuntz]{Timothy Clifton\thanks{E-mail:
  tclifton@astro.ox.ac.uk} and Joe Zuntz\\ 
Oxford Astrophysics, Physics, DWB, Keble Road, Oxford, OX1 3RH, UK}
\begin{document}


\maketitle
\label{firstpage}

\begin{abstract}
We consider the effects of large structures in the Universe on the Hubble
diagram.  This problem is treated non-linearly by considering a
Swiss Cheese model of the Universe in which under-dense voids are
represented as negatively curved regions of space-time.  Exact
expressions for luminosity distances and redshifts are used to
investigate the non-linear effects of structure on the magnitudes of
astrophysical sources.  It is found that the intervening voids we
consider, between the observer and source, produce changes in apparent magnitude of less
than $0.012$.  Sources inside voids, however, can be affected
considerably at redshifts below $z \sim 0.5$.  By averaging observable
quantities over many randomly generated distributions of voids we find
that the presence of these structures has the effect of introducing a
dispersion around the mean, which itself can be displaced the
background value.  Observers in an inhomogeneous universe,
who take averages of observables along many different lines of sight,
may then introduce systematic biases, and under-estimate errors, if
these effects are not taken into account.  Estimates of the potential size
of these effects are made using data from simulated large-scale
structure.
\end{abstract}



\section{Introduction}

Hubble diagrams play a key role in our understanding of the evolution
of the Universe.  It was Hubble diagrams that first led
to widespread acknowledgement of the expanding Universe paradigm, and
today, in the form of type Ia supernova observations, they provide
important evidence for the Dark Energy that is at the heart of the $\Lambda$CDM
model of the Universe. Ongoing and future projects aim
to collect more and more data in order to reconstruct the expansion history of
the Universe to ever increasing accuracy, and to test hypotheses about the nature
of Dark Energy itself.

Given the extraordinary implications of the supernova results, and
the large amounts of resources that are being invested in them, it
seems prudent to make sure we fully understand all physical effects 
that may bias, or influence, the conclusions which are
drawn from them.  To this end, we perform a detailed, and fully non-linear,
investigation of the effects of the simplest large structures on Hubble diagrams.

That structure exists on small scales in the Universe is, of course,
indisputable, but while some studies of galaxy surveys have pointed toward homogeneity on scales of $\sim
70 h^{-1}$Mpc \citep{Hogg}, others have concluded that the largest structures so far detected are limited only
by the size of the surveys that found them \citep{Labini,Labini2,Labini3}.  The apparent recent
detection of an anomalously large local bulk flow \citep{bulk}, as well
as the existence of unexpected features in the CMB \citep{evil,cold}, and
the CMB dipole itself, also hint at the possibility of large structures existing
in the Universe.  Here we do not wish to debate the evidence for or
against structure existing on different scales, but rather to
calculate the effects that different structures have on Hubble
diagrams.

The majority of studies in this area have been performed within the
context of linear perturbation theory
\citep{linear1,linear2,linear3,linear4,linear5,linear6,weak,linear7,linear8,linear9,linear10,weak2}.
Here we treat the problem non-perturbatively by modelling the Universe as a
Friedmann-Robertson-Walker (FRW) background, with spherical sections removed and
replaced by regions of Lemaitre-Tolman-Bondi (LTB) space-time
\citep{LTB1,LTB2,LTB3}. With an
appropriate choice of boundary conditions between the FRW and LTB
regions, the resulting geometry is an exact solution of Einstein's
equations.  Such a solution is often referred to as a Swiss Cheese
universe, although the replaced regions here are not completely
empty.  This method should be considered complimentary to studies
performed in the linear frame-work:  Both require some degree of
approximation, but different types of approximation in each case.

The Swiss Cheese approach has a number of drawbacks, but its great
benefit is that it allows calculations to be performed beyond the
linear level.  As the solution is exact, and as it is possible to derive
simple expressions for redshift and luminosity distances within it, all
higher order and non-perturbative effects are automatically included.
Luminosity distances in LTB Swiss Cheese universes have been studied
previously by \citet{biswas, onion, strong, strong2} and
\citet{marra}. We will compare our results with these
studies as we proceed.  

It has been suggested in the literature that non-linear effects from
large-scale inhomogeneities may be responsible for the apparent
detection of Dark Energy (see, for example, \citet{mattsson}).
Our goal here is not to construct a situation in which one can fit for the observations
without Dark Energy, but rather to attempt to
calculate the effect of inhomogeneities on Hubble diagrams for the
purposes of better understanding their influence on parameter estimation, or,
conversely, on constraining structure, if cosmological parameters are
deemed to be known from elsewhere. Our results indicate that the type
of structures expected to exist in the Universe are highly unlikely to be
capable of successfully mimicking Dark Energy.

In section \ref{theory} we introduce the theory.  We will briefly
discuss the LTB solution, and how it can be embedded into an FRW
universe.  We then go on to a more detailed discussion of redshifts
and luminosity distances in the resulting space-time.  This is done by
considering bundles of null geodesics, and the Sachs optical
equations.  

In section \ref{singlevoid} we investigate the effect of a single
large void in the Universe on luminosity distances as functions of
redshift.  Measures of distance to objects on the other side of the
void appear largely unaffected by its presence, with changes in
apparent magnitude of less than $0.012$, for the voids we consider.  This is consistent with
linear treatments, such as that of \citet{weak}, as well as previous non-linear
treatments, such as that of \citet{strong, strong2}. Observations of objects that
reside inside the void, however, can be considerably affected by the
void's presence, with shifts in apparent magnitude of up to $0.2$ being
easily obtainable at low redshifts ($z < 0.2$).  At higher
redshifts the effect on objects inside the void drops off, and becomes
sub-dominant compared to the small effect of looking through them.
We present results for the change in luminosity
distance that can result from voids of varying depths, widths and at
different redshifts.  Results here are limited to the case of looking
through the centre of voids.

Section \ref{manyvoids} contains an analysis of the effect
of looking through many voids in a row.  In this section the voids are
drawn from idealised distributions.  Examples are presented, and the
case of averaging over many lines of sight is analysed.  Such
averaging of luminosity distances, for specified geometries, appears
to us to be a preferable, if more cumbersome method, than averaging
the mass distribution, and calculating a single luminosity distance in
the corresponding geometry.  We present results for the dispersion,
and deviation, of Hubble diagrams that results from different
distributions of voids.  As the considerations of single voids
suggests, the effect of objects located inside voids contributes most
of the dispersion at low redshifts, while at high redshifts the
dispersion is mainly due to the cumulative effect of looking through voids.

In section \ref{realvoids} we make an attempt at linking the idealised
cosmologies, considered in previous sections, to some more realistic
distributions of matter.  The idea here is that instead of taking
purely idealised distributions of voids, we take density profiles from simulated
large-scale structure, and then use these distributions to motivate
our Swiss Cheese models.  To achieve this we consider simulated
structure generated from the Millennium Simulation.  These profiles are produced by a process of averaging over
different length scales. The profiles obtained are then idealised, so as
to fit into the framework developed in the preceding sections, and the
results on Hubble diagrams are calculated.  This method is not
proposed as a way of superseding the linear treatment of distance measures
within these space-times (which would manifestly be applicable here, as
they are created in the linear regime).  Rather, we intend it to be a
method of obtaining realistic distributions of voids from a well motivated
source. 

Finally, in section \ref{conclusions}, we conclude.  The Appendix
shows an interesting example of another case that can be solved for
exactly.

\section{Swiss Cheese Cosmology}
\label{theory}

The cosmology we consider here is an FRW background with spherical
regions removed and replaced with LTB space-times, in order to model
inhomogeneities. Sub-sections \ref{A2}-\ref{C} recap results on LTB
space-times, boundary conditions for embedding LTB patches into FRW,
and expressions for null geodesics and redshifts in LTB. Sub-section
\ref{D} then contains a discussion of luminosity distances in these space-times, for
both observer and source away from the centre of symmetry. The FRW
and observer centred limits of these expressions are found in
sub-section \ref{E}.

\subsection{The LTB Solution}
\label{A2}

The LTB line-element is given by \citep{LTB1,LTB2,LTB3}
\be
\label{ltb}
ds^2=-dt^2 + \frac{R^{\prime 2}}{(1-k r^2)}dr^2 +R^2 d\Omega^2,
\ee
where $R=R(t,r)$, $k=k(r)$ and prime denotes partial differentiation
with respect to $r$. For $\Lambda=0$ and $k<0$ we can write $R$ in parametric form as
\bea
\label{eds}
R &=& \frac{m (1-\cosh \Theta )}{2 k r^2}\\
t-t_0 &=& \frac{m (\sinh 2 \Theta-2 \Theta )}{2 (-k r^2)^{3/2}}
\label{eds2}
\eea
where $t_0=t_0(r)$ and $m=m(r)$.  Exact solutions exist for
$\Lambda \neq 0$, and are given in terms of elliptic functions by
\citet{lambda}.  It is also possible to solve for the case with $k \geq
0$, but this will not be required here. The geometry (\ref{ltb}) is an exact solution of Einstein's equations
in the presence of a perfect fluid of pressureless dust with
energy density
\be
\label{rho}
\rho = \frac{m^{\prime}}{R^2 R^{\prime}}.
\ee
Gauge freedoms allow us to transform into a coordinate system in
which $t_0=$ constant, without loss of generality. The LTB space-time
is then completely specified by a choice of $k(r)$ and $m(r)$, and
reduces to FRW in the limit $k(r)$ and $m(r)=$constant.
We will refer to $k(r)$ as the spatial curvature and $m(r)$ as the
gravitational mass distribution.

\subsection{Boundary Conditions}
\label{B}

We now wish to replace regions of FRW with the LTB
geometry described above. To do this we need the conditions required to match a
manifold with metric (\ref{ltb}) to one with FRW metric
\be
ds^2 = -dt^2 +a^2(t) \left( dx^2 +f^2(x) d\Omega^2 \right)
\ee
at a boundary of constant $r=x=\Sigma$.  The Darmois junction conditions imply
that the matching is a solution of Einstein's equations if the first
and second fundamental form on the hyper-surface at $\Sigma$ are
identical on either side \citep{boundary}.  This
corresponds to the two conditions \citep{ribeiro}
\bea
\label{boundary1}
R \vert_{\Sigma} &= a f \vert_{\Sigma}\\
\sqrt{1- k r^2}\vert_{\Sigma} &= f^{\prime} \vert_{\Sigma }.
\label{boundary2}
\eea
It can be shown that equations (\ref{boundary1}) and (\ref{boundary2})
imply that the `Bondi mass' inside the excised region should equal the
mass of the FRW region that was removed \citep{ribeiro}:
\bea
\nonumber
m^{(LTB)}&=&4 \pi \int \rho R^{\prime} R^2 dr \\ &=&4 \pi \int \rho a^3
f^2 f^{\prime} dx = 4 \pi \rho a^3 f^3 = m^{(FRW)},
\label{boundary3}
\eea
where $m^{(LTB)}$ in the mass from Eq. (\ref{rho}).  These conditions
can be verified to describe the requirement that the LTB metric
should reduce to FRW at the boundary.  

\subsection{Null Geodesics and Redshifts}
\label{C}

Now consider a null geodesic in (\ref{ltb}) with affine parameter $\lambda$, and with
a tangent vector $k^a = d x^a/d \lambda$.  If $n^a$ is the unit
vector in the direction of $k^a$, in the rest space of an observer with
4-velocity $u^a$, then $k^a$ can be decomposed as
\begin{equation}
\label{decomp}
k^a = (-u_b k^b) (u^a +n^a),
\end{equation}
where $u_a u^a=-1$, $n_a n^a = 1$ and $u_a n^a=0$.  Thus, for a radial
null geodesic, and an observer co-moving with the coordinate system
(\ref{ltb}), we have $u^a=(1,0,0,0)$ and $n^a=(0,\pm
\sqrt{1-kr^2}/R^{\prime},0,0)$.  The $\pm$ sign here corresponds to
geodesics directed away, or toward, the centre of symmetry, and we
have chosen $\lambda$ to increase with $t$.
According to (\ref{decomp}), an infinitesimal increment in affine parameter, $d\lambda$, will
then be seen by the observer following $u^a$ as changes in time and
position of
\begin{eqnarray}
\label{dt}
dt&=&-u_a k^a d\lambda\\
\label{dr}
dr&=&\mp u_a k^a  \frac{\sqrt{1-kr^2}}{R^{\prime}} d\lambda.
\end{eqnarray}
This agrees with the equation for radial null geodesics, which can be read off from (\ref{ltb}) as
\begin{equation}
\label{r}
\frac{dr}{dt} = \pm \frac{\sqrt{1-kr^2}}{R^{\prime}}.
\end{equation}
The tangent vectors for radial geodesics can now be written as $k^a =
(A,B,0,0)$, where $A=A(t,r)$ and $B=B(t,r)$.  The trajectories to
which $k^a$ are tangent are both null and geodesic, so that $k^a k_a=0$ and
$k^a {k^b}_{;a} =0$.  This gives $B=\pm A \sqrt{1-kr^2}/R^{\prime}$, and
\begin{equation}
\label{k}
k^a =A\left( 1,\pm \frac{\sqrt{1-kr^2}}{R^{\prime}} ,0,0 \right),
\end{equation}
where $A$ as the solution of
\begin{equation}
\label{A}
\pm \sqrt{1-kr^2} A^{\prime} + \dot{A} R^{\prime} + A
\dot{R}^{\prime}=\frac{dA}{dt} R^{\prime} + A \dot{R}^{\prime}= 0.
\end{equation}                   
This expression integrates to
\begin{equation}
\label{A3}
A \propto \exp \left\{ - \int
\frac{\dot{R}^{\prime}}{R^{\prime}} dt \right\}.
\end{equation}
As always, the redshift of a photon is given by
\begin{equation}
1+z = \frac{(u^a k_a)_e}{(u^b k_b)_o},
\end{equation}
where subscript $e$ denotes a quantity at the point where the photon is
emitted, and subscript $o$ a quantity at the point it is observed.  In
the present situation we have $u^a k_a = -A$, so the redshift is
\begin{equation}
\label{z}
1+z = \frac{A_e}{A_o} = \exp \left\{ \int^o_e \frac{
  \dot{R}^{\prime}}{R^{\prime}} dt \right\},
\end{equation}
in agreement with the source centred case considered by
\citet{LTB3}, and the observer centric case considered by
\citet{ribeiro}. This expression is valid for any source and observer
connected by a radial geodesic, if they are centred or not.

\subsection{Luminosity Distances}
\label{D}

Consider a bundle of null geodesics with cross-sectional area $d\sigma$,
and tangent vector $k^a$.  The rate of change of
$d\sigma$ along the bundle is
\begin{equation}
\label{S}
\frac{d (d \sigma)}{d\lambda} = {k^a}_{;a} d\sigma,
\end{equation}
and is independent of the 4-velocity of the screen onto which it is
projected \citep{Sachs}.  Differentiating this expression with respect
to $\lambda$ gives the second order Sachs optical equation:
\begin{equation}
\label{secondorder}
\frac{1}{\sqrt{d \sigma}}  \frac{d^2 \sqrt{d \sigma}}{d \lambda^2} = -
\left( \vert \varsigma \vert^2 + \omega^2 + \frac{1}{2} R_{a b} k^a k^b \right),
\end{equation}
where $\vert \varsigma \vert^2= k_{(a;b)}k^{a;b}/2-({k^a}_{;a})^2/4$ is
the shear, and $\omega^2= k_{[a;b]}k^{a;b}/2$ is the rotation of the
bundle.  For radial geodesics we then have $\varsigma=0$, from
symmetry considerations, and we can set $\omega=0$, as the sources
under consideration are effectively point-like.

For the radial geodesics (\ref{k}), in the space-time (\ref{ltb}), the
right hand side of equation (\ref{secondorder}) can be shown to be
given by \citep{onion}
\begin{equation}
\label{secondorder2}
\frac{1}{\sqrt{d \sigma}}  \frac{d^2 \sqrt{d \sigma}}{d \lambda^2} =
\frac{1}{R}  \frac{d^2 R}{d \lambda^2},
\end{equation}
which can be verified using (\ref{dt}), (\ref{dr}) and (\ref{A}).  The
solution to this equation is given by
\begin{equation}
\label{dsig}
\sqrt{d \sigma} \propto R \int \frac{d \lambda}{R^2},
\end{equation}
where the constant of proportionality can be found by considering the
limit $d\sigma \rightarrow 0$, when the beam is focused at the
observer.  We then have
\begin{equation}
\frac{d \sqrt{d \sigma}}{d \lambda} \rightarrow d \Omega,
\end{equation}
where $d\Omega$ is the square root of the solid angle subtended by the
beam.  Enforcing this limit in equation (\ref{dsig}) gives
\begin{equation}
\sqrt{d \sigma} = d\Omega R_o R \int \frac{d \lambda}{R^2},
\end{equation}
where $R_o$ is the value of $R$ at the observer, where the beam is
focused.

The angular diameter distance along a radial geodesic in this
space-time is then given by
\begin{equation}
\label{ang}
r_A \equiv \frac{\sqrt{d \sigma_e}}{d\Omega} = R_o R_e \int_o^e \frac{d
  \lambda}{R^2},
\end{equation}
which is the expression obtained by \citet{onion}.
The galactic angular distance, along the same geodesic, is then given
by Etherington's theorem \citep{etherington} as
\begin{equation}
r_G= (1+z) R_o R_e \int_o^e \frac{d \lambda}{R^2},
\end{equation}
and the luminosity distance is
\begin{equation}
\label{lum}
r_L=(1+z)^2 R_o R_e \int_o^e \frac{d \lambda}{R^2}.
\end{equation}
This is the required expression for luminosity distance, applicable
when both source and observer are off centre.

\subsection{FRW and Observer-Centric Limits}
\label{E}

To clarify how the redshifts and distance measures described above
relate to more usual expressions we will now consider the FRW, and observer
centred LTB limits of these equations.

In the FRW limit of LTB space-times we have that $k(r)$ and $m(r)\rightarrow$constant
and, consequently, $R(t,r) \rightarrow a(t) r$.  In this limit the LTB line-element
(\ref{ltb}) reduces to its FRW counterpart.  It can also be seen
that the expression for redshift, equation (\ref{z}), reduces to
\begin{equation}
1+z \rightarrow \exp \left\{ \int^o_e \frac{\dot{a}}{a} dt \right\} =\frac{a_o}{a_e},
\end{equation}
which is, of course, the usual FRW expression.

If we now consider the angular diameter distance, (\ref{ang}), we can
see that, using (\ref{dr}) and (\ref{A}), this reduces to
\begin{equation}
r_A \rightarrow a_0 a_e r_o r_e \int_o^e \frac{dr}{r^2 \sqrt{1- k r^2}} =
a_o a_e r_o r_e \left[ \frac{\sqrt{1-k r^2}}{r} \right]^o_e,
\end{equation}
as, in this limit, we have $A \propto 1/a$.  To confirm that this is
the usual FRW angular diameter distance it is convenient to transform
to the proper radial coordinate, $l$, via $r = f(l)$,
where:
\begin{equation}
  \begin{array}{ l l l }
     f(x)=\sin(x)  & \; & k=1\\
     f(x)=x & \qquad \mbox{for} \qquad & k=0\\
     f(x)=\sinh(x)  & \; & k=-1.
  \end{array}
\end{equation}
The FRW limit of the angular diameter distance can then be written in
the more familiar form
\begin{equation}
r_A \rightarrow \frac{f(\Delta l)}{(1+z)},
\end{equation}
where $\Delta l= \vert l_e-l_o \vert$.  The Minkowski limit follows
from this, with $a=$constant and $k=0$, and is given by $r_A
\rightarrow \Delta r$.

Finally, let us consider the observer centric case, which should
follow from the more general expressions above when $R_o \rightarrow
0$.  The most convenient way to find this limit is to differentiate
(\ref{ang}), with respect to $\lambda$, to obtain
\begin{equation}
\frac{1}{R_e} \frac{d r_A}{d\lambda} - \frac{r_A}{R_e^2} \frac{d
  R_e}{d\lambda} = \frac{R_o}{R_e^2} \rightarrow 0.
\end{equation}
Integrating the left-hand side of this then gives
\begin{equation}
\label{rL}
r_A \propto R_e,
\end{equation}
which is the expected form of the angular diameter distance for an
observer at the centre of symmetry \citep{LTB3}.  The expression for
redshift, (\ref{z}), can immediately be seen to reduce to the observer
centred case considered by \citet{ribeiro}, as $r_o \rightarrow 0$.

The expressions for redshift and luminosity distance, (\ref{z}) and
(\ref{lum}), therefore have the correct FRW, Minkowski and
observer centred limits, and reduce to the known expressions as they
are approached.

\section{A Single Void}
\label{singlevoid}

We will now consider the case of a flat FRW Universe containing a
single void.  We can choose coordinates, without loss of generality,
such that $t_0=$constant, and so that the big bang occurs simultaneously
at all points in space.  This condition is automatically satisfied in
FRW if surfaces of constant density are chosen to be surfaces of equal
time.  We also choose initial conditions such that $m
\propto r^3$, so that the gravitational mass is initially evenly
distributed.  The form of the void is then specified by the function
$k(r)$, which we will take to be a smooth curve of the form
\begin{equation}
\label{kr}
k=\frac{k_0}{2} \left[ 1+  \cos \left(\pi \frac{r}{r_0} \right) \right],
\end{equation}
where $k_0<0$ and $r_0>0$ are constants specifying the depth and width
of the void, respectively.  This negative perturbation in $k$ is a
smoothly varying function of $r$, that reduces to $0$ at the edge of
the void, $r=r_0$.  The negative curvature causes
the space inside the void to expand faster than that
outside, so that the energy density inside the void is dissipated more
rapidly.  An under-density then occurs, not due to any any decrease in
gravitational mass, but due to an increase in spatial volume.
Although the choices above appear reasonably natural, one may consider
more general space-times, either by changing the functional form of
$m(r)$ and $k(r)$, or by considering less symmetric space-times than
that of LTB.  We will postpone considering the effects of such
generalisations here, so that we can first concentrate on
the simple case of the smooth perturbation in curvature outlined above.

\subsection{An Einstein-de Sitter background}

\begin{figure}
\begin{flushright}
\vspace{-15pt}
\epsfig{figure=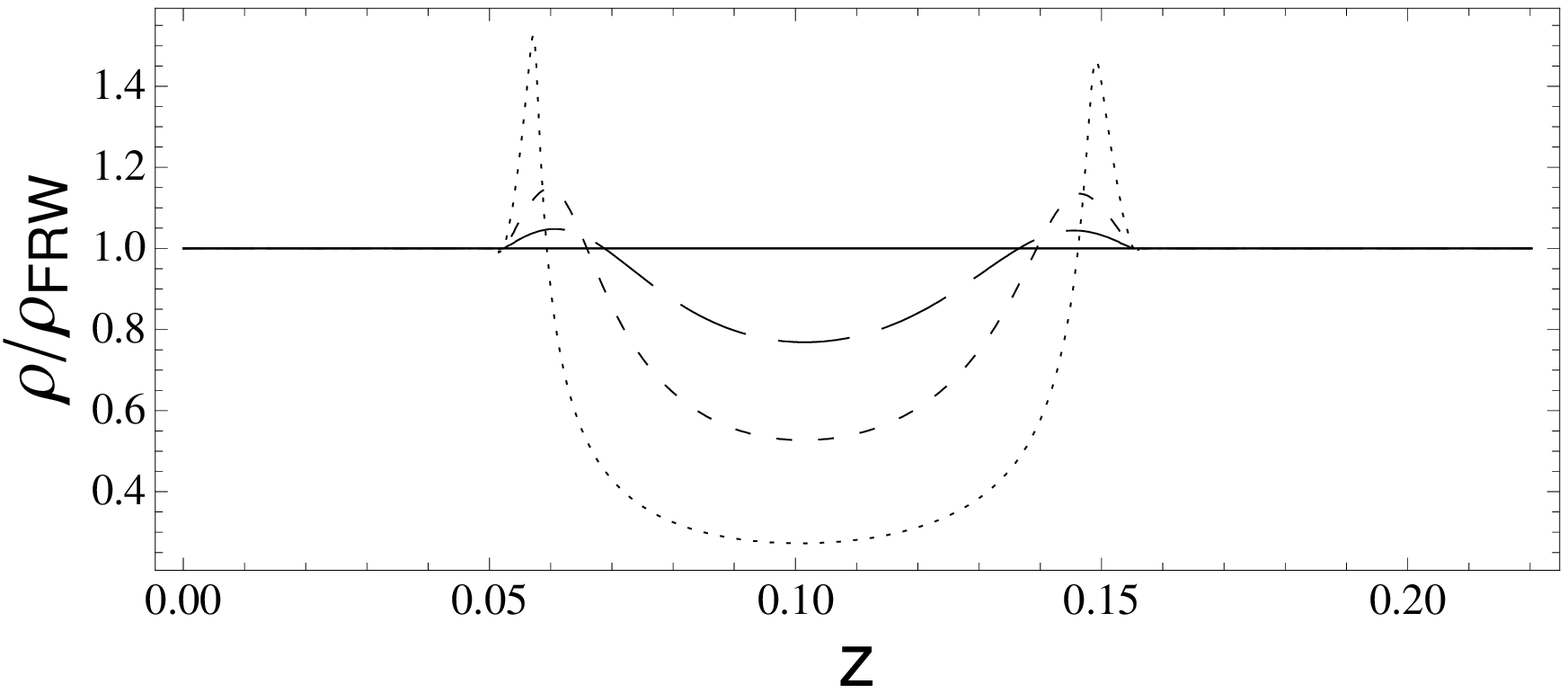,width=8.35cm}
\epsfig{figure=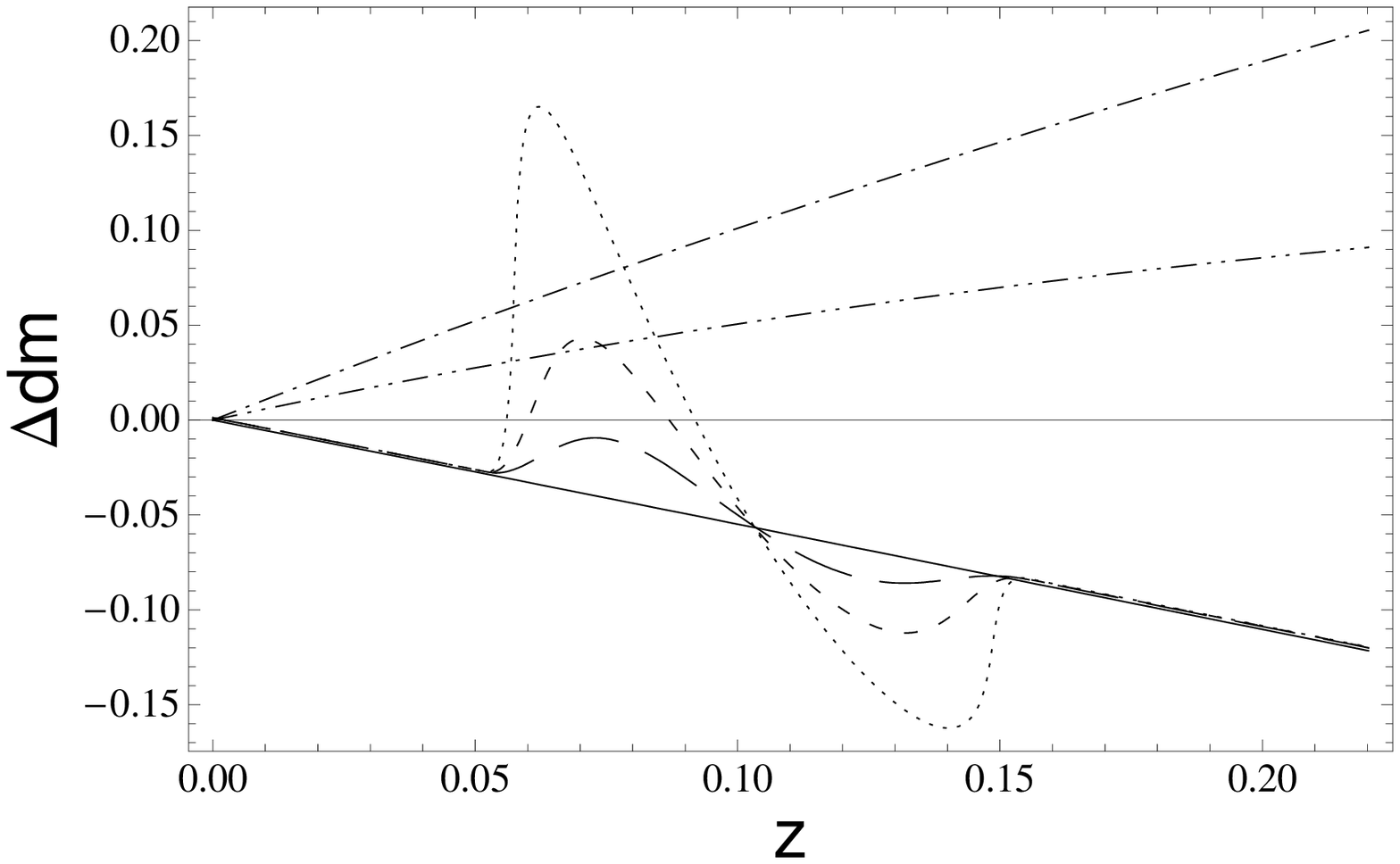,width=8.55cm}
\end{flushright}
\vspace{-10pt}
\caption{The upper panel shows the fractional energy density experienced by the
  photon, where the long-dashed, short-dashed and dotted lines
  correspond to voids that are $25\%$, $50\%$ and $75\%$ under-dense today. The lower
  panel show how these three voids effect the distance modulus.  The
  solid, dot-dashed and double-dot-dashed lines in this plot correspond
  to EdS, dS and $\Lambda$CDM with $\Omega_{\Lambda}=0.7$,
  respectively.  These results are with an EdS background.}
\label{fig1}
\end{figure}

Consider the case of a void in a space-time that is
asymptotically a spatially flat Einstein-de Sitter (EdS) universe, as $r \rightarrow r_0$ in
(\ref{ltb}).  In this case $\Lambda=0$, and the metric functional
$R(t,r)$ is given parametrically by equations (\ref{eds}) and
(\ref{eds2}).  The functions $t=t(z)$ and $r=r(z)$, along past null
geodesics, can then be found by integrating equations (\ref{r}) and
(\ref{z}). Substituting these expressions into equation (\ref{lum})
gives $r_L(z)$. Here we will use this result to investigate the effects such 
structures have on Hubble diagrams, when observers look through them.

We begin by considering the effect of voids with varying width and
depth, at the same position.  For ease of computation we consider the
space-time as an LTB under-density matched to an FRW background under
the conditions (\ref{boundary1}) and (\ref{boundary2}), at $r=r_0$.
The geodesic equations, and redshift relations, must
now be integrated in each space-time and $r(z)$ and $t(z)$ matched at
the boundaries.  The cumulative redshift, $1+z_T$, is given in the usual
way by the expression $$(1+z_T) = \prod_i (1+z_i),$$ where $1+z_i$
is the redshift along a portion, $i$, of the geodesic.

The effects of three voids with different depths, all approximately
the same width and located at the same redshift ($z \sim
0.1$) are shown in Figure \ref{fig1}.  The upper plot shows the energy
density encountered by a photon as it travels through the three
different voids, and the lower plot shows the corresponding distance
modulus, $\Delta$dm, as a function of redshift, $z$.  Distance
modulus is defined as the magnitude an object appears at, minus the
magnitude it would have at the same redshift in a empty, negatively curved Milne universe.
It can be written in terms of luminosity distance as
\be
\Delta \textrm{dm} = 5 \log_{10} r_L-5 \log_{10} r_L^m,
\ee
where $r_L$ is given by (\ref{lum}), and $r_L^m=z+z^2/2$ is the
expansion normalised luminosity distance in a Milne universe.
Anything below $\Delta$dm=$0$ in Figure \ref{fig1} is then interpreted
as a decelerating universe, and anything above it as accelerating.
For reference, we have included in this plot the distance moduli for
an EdS universe, as the solid line, and for a de Sitter (dS)
space, as the dot-dashed line.  A $\Lambda$CDM cosmology, with
$\Omega_{\Lambda}=0.7$, is shown as the double-dot-dashed line.

\begin{figure}
\begin{flushright}
\vspace{-15pt}
\epsfig{figure=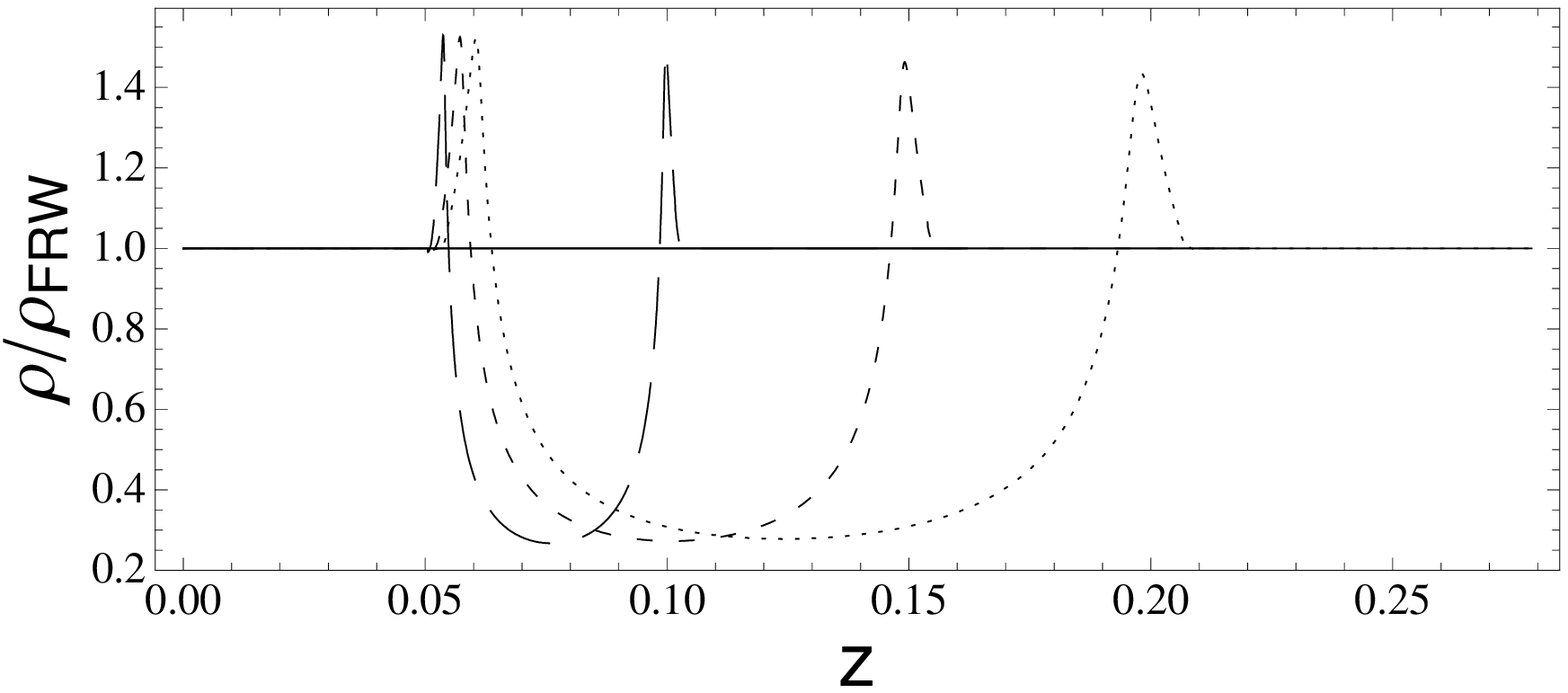,width=8.35cm}
\epsfig{figure=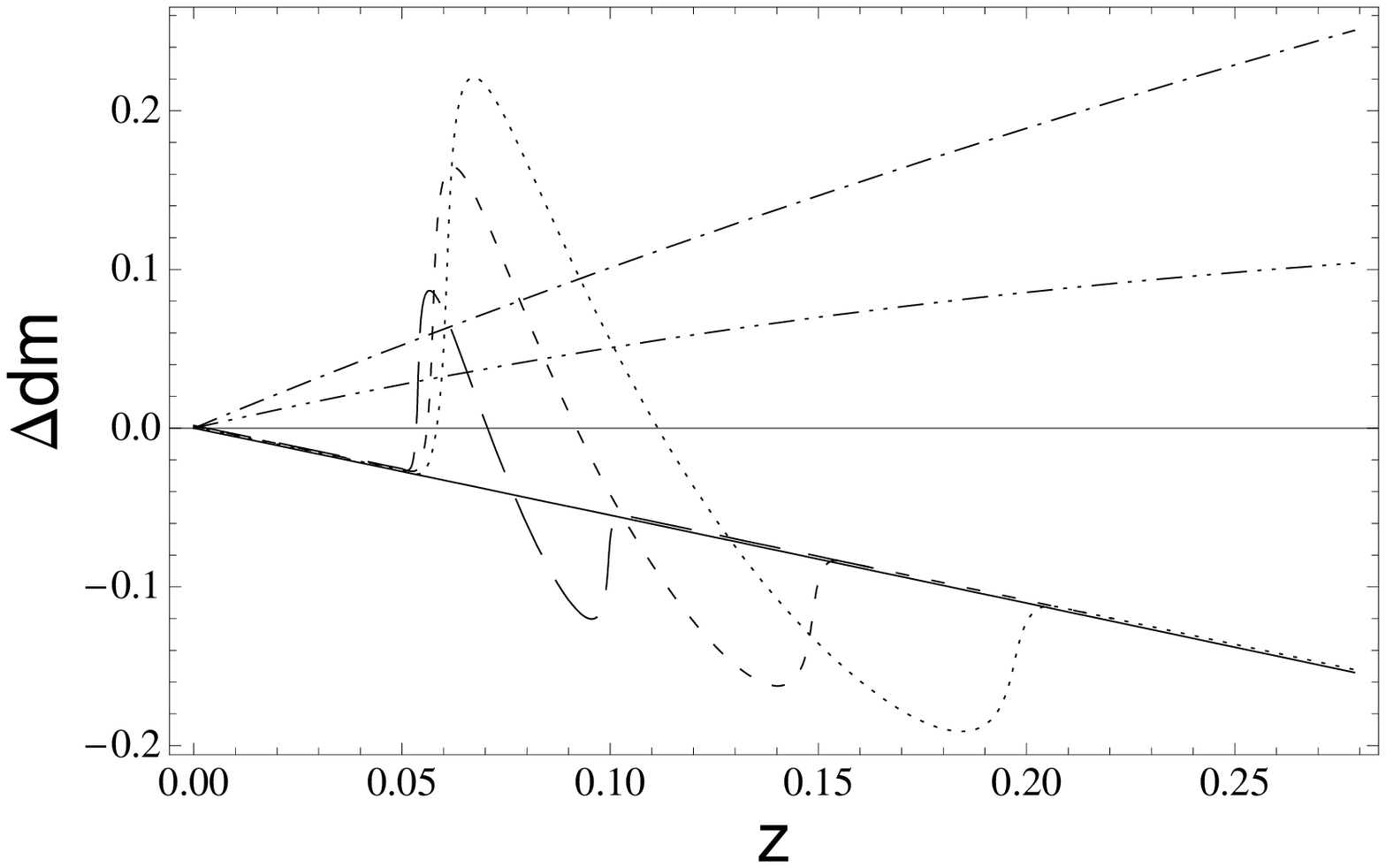,width=8.55cm}
\end{flushright}
\vspace{-10pt}
\caption{The upper panel shows the fractional energy density for three
voids of the same depth, but with varying widths.  The lower panel
shows the corresponding effect on the distance modulus.  Dot-dashed,
double-dot-dashed and solid lines are as in Figure \ref{fig1}. These
results are with an EdS background.}
\label{fig2}
\end{figure}

It can be seen from the upper plot of Figure \ref{fig1} that, although
the spatial curvature $k$ is always negative, the energy density can
reach values in excess of the asymptotic value at the edge of the
void.  This is due to the analogue of the radial scale factor,
$R^{\prime}$, being lower than its asymptotic value in this region.  Aside from these small
over-densities, at the edges of the void, it can clearly be seen that
the energy density in the centre of the void has been dissipated as a result
of the more rapid expansion there, caused by the negative curvature
perturbation.  The three voids shown in this plot are chosen such that
at the their centres they are $25\%$, $50\%$ and $75\%$ under-dense at
the present time.  The upper plot shows these regions to be slightly less
under-dense than this at their minima, as what is depicted is the
energy density experienced by the photon as it passed through them,
sometime before the present.

The lower plot in Figure \ref{fig1} shows that there is no noticeable
effect, due to the void, on viewing objects that are beyond it.  While the
presence of the voids causes a deviation from the EdS background in their vicinity, it can
be seen their distance moduli return to within $0.01$ of the background
value at redshifts beyond.  Weak lensing effects are expected to be
within this order of magnitude (see e.g. \citet{weak}) as well as the
late-time integrated Sachs-Wolfe effect (see e.g. \citet{ISW}).
However, it seems unlikely that such small effects
will be detectable in Hubble diagrams that are constructed from supernovae
observations, at least in the foreseeable
future\footnote{The `intrinsic
  errors' in the magnitude of supernovae are currently of the order
  $0.1-0.2$.}.  Hence, the principle effect of these structures appears to be
on objects that are inside the voids themselves, in which case the deviation from the background
value can be seen to be considerable for the examples shown here.
Clearly these deviations are a function of the void depth, and
increase in a proportionate way.  We note this effect is appears to be
largely due to the $0.1-0.3$ perturbations that appear in the metric functional
$R(t,r)$.  Such large metric perturbations are unlikely to develop in
the linear approach, and so this effect may not show up in such a
pronounced way in treatments of that kind.

\begin{figure}
\begin{flushright}
\vspace{-15pt}
\epsfig{figure=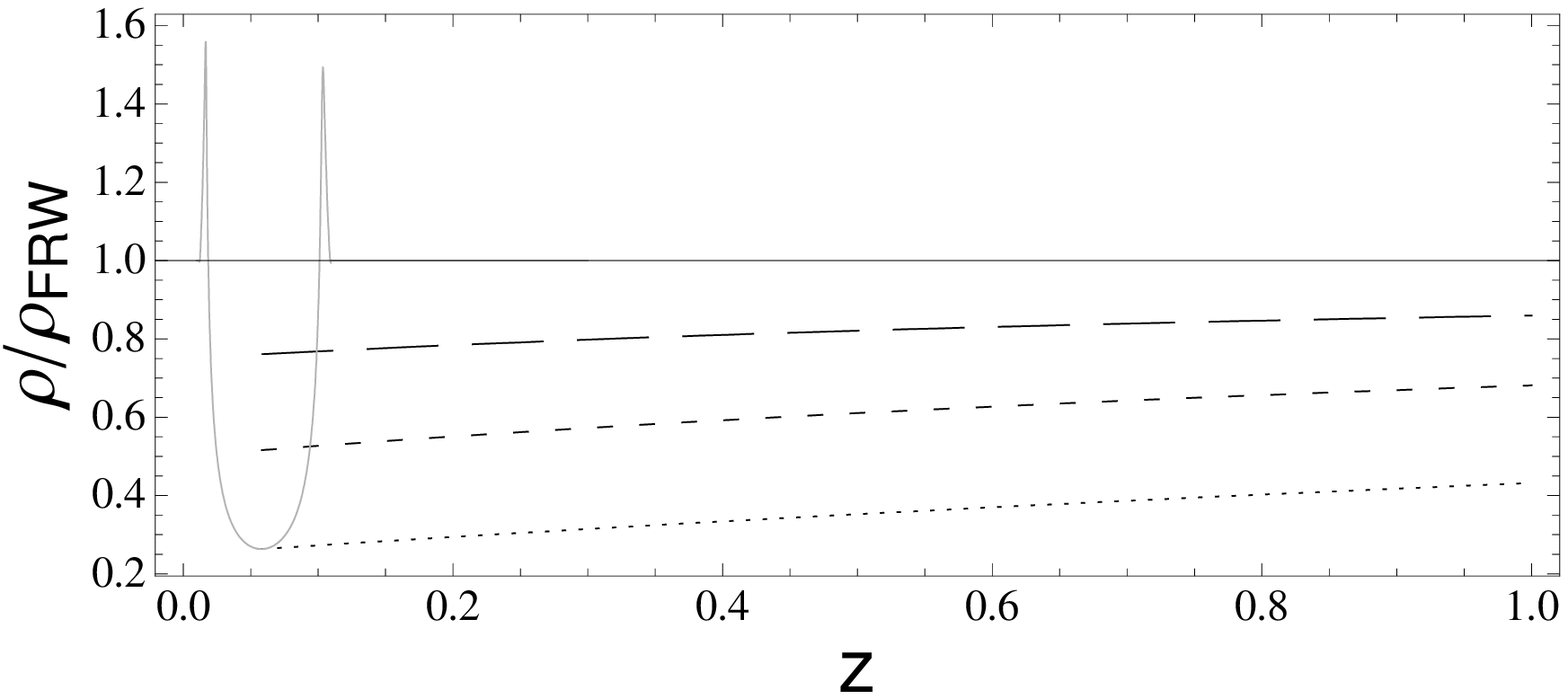,width=8.35cm}
\epsfig{figure=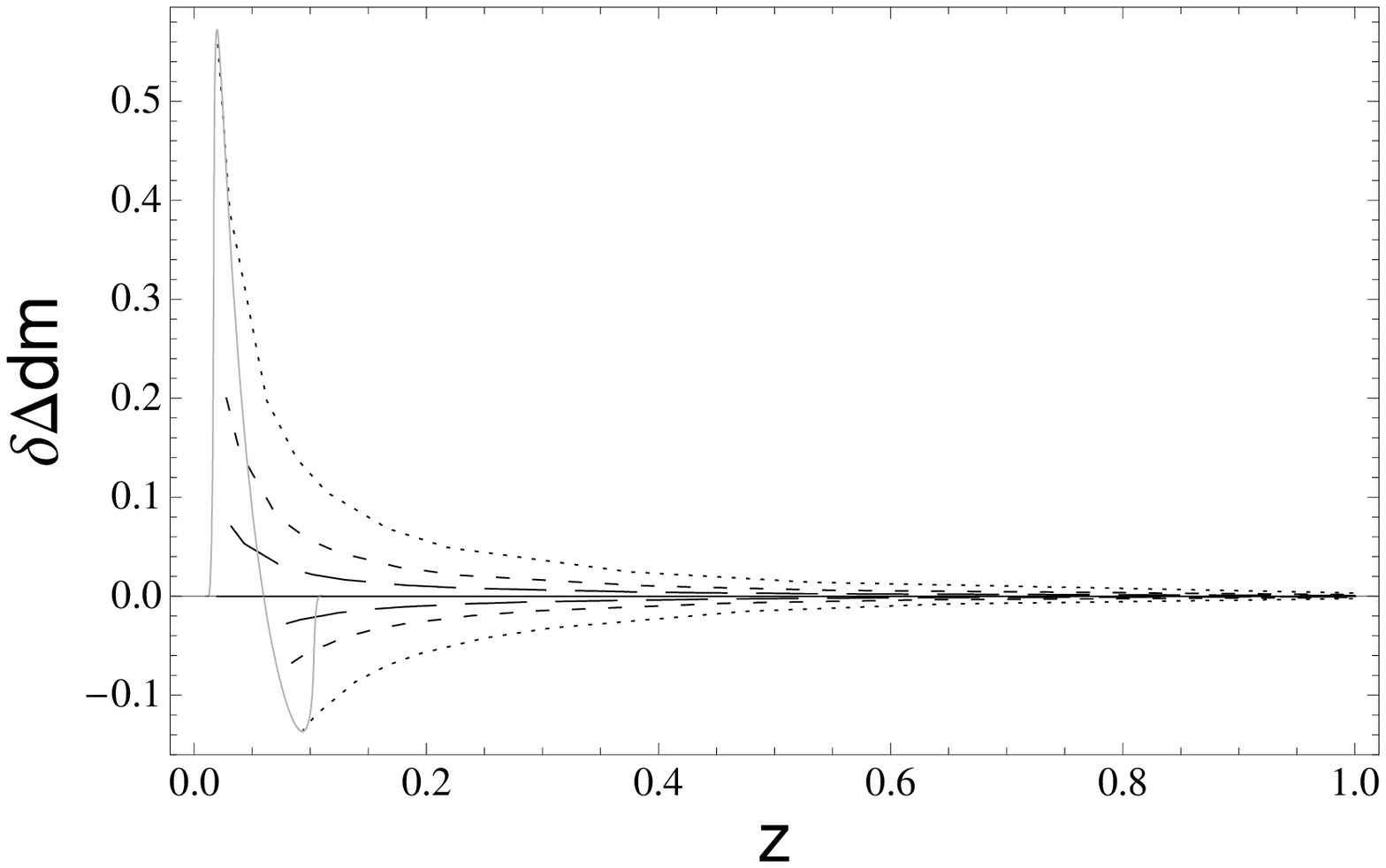,width=8.6cm}
\end{flushright}
\vspace{-10pt}
\caption{The effect of changing the location
  of the void.  The three voids from Figure \ref{fig1} are moved to
  different redshift between $z=0.1$ and $1$.  The upper panel shows
  how the fractional energy density at the centre of the voids varies
  with redshift.  The lower panel shows the maximum displacement in
  distance modulus, from the background value, that these voids
  cause. These results are with an EdS background.}
\label{fig3}
\end{figure}

The small effect of the voids on viewing objects behind them is in
accordance with the studies of \citet{strong, strong2} and \citet{biswas}, who
found similar results in their studies of luminosity distances in
Swiss cheese.  The maximum shift in apparent magnitude for objects
also seems consistent with the results of \citet{biswas}, who find the
effect of a void with a $50\%$ under-density at its centre produces a
change in apparent magnitude of $\delta \Delta$dm$\sim 0.4$.  This can
be seen to be similar to that found in Figure \ref{fig1}.

As well as varying the depth of a void, we will also be interested in
the effect of varying its width, and its distance from us.
This is shown in Figures \ref{fig2} and \ref{fig3}.  In Figure
\ref{fig2} all three example voids are now chosen to have the same
depth, so that they are $75\%$ under-dense today, while their width is
varied.  The deviation from the background distance modulus changes
its width accordingly, and as would be expected.  We note that the
maximum deviation is larger for the widest voids, decreasing
substantially as the width of the void is decreased.

\begin{figure}
\begin{flushright}
\vspace{-15pt}
\epsfig{figure=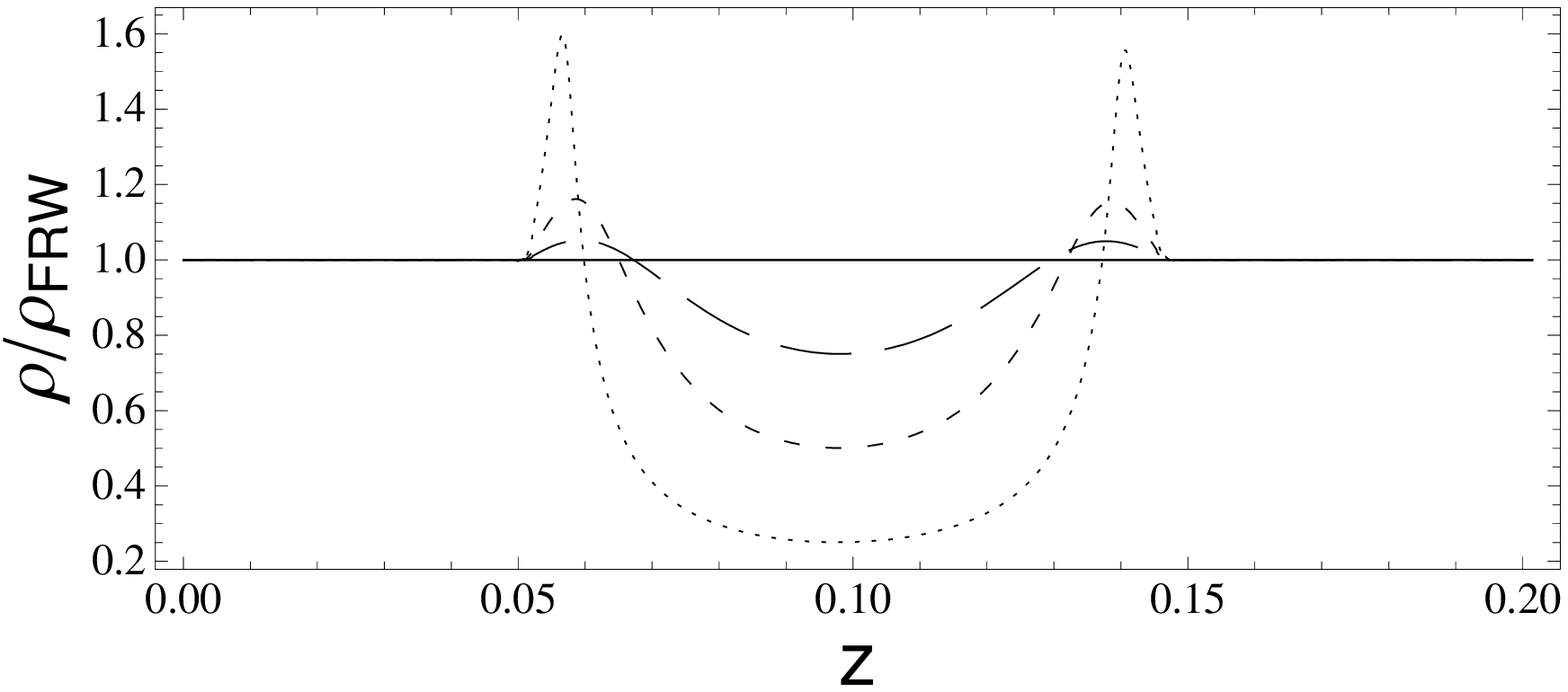,width=8.2cm}
\epsfig{figure=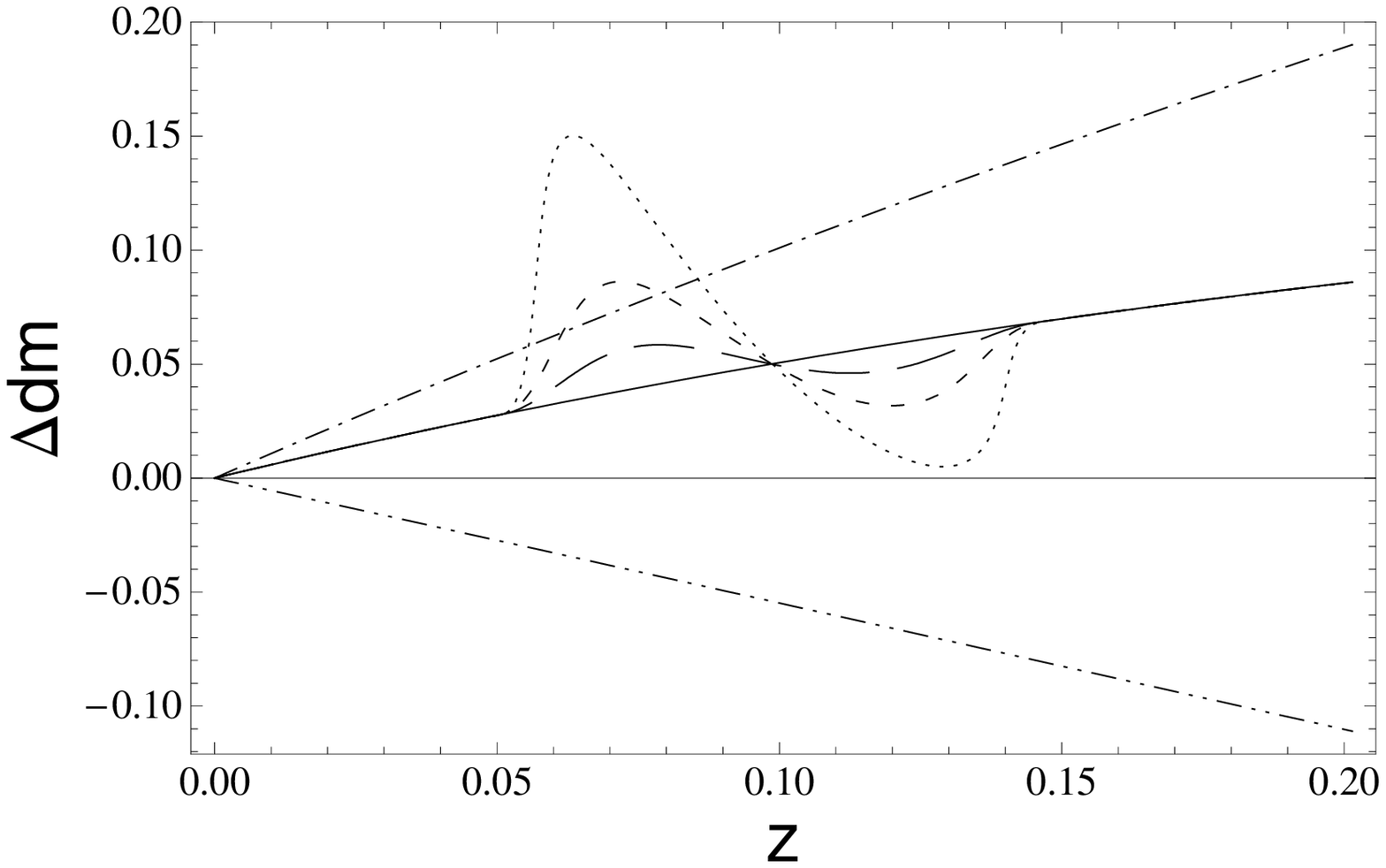,width=8.6cm}
\end{flushright}
\vspace{-10pt}
\caption{The upper panel is the same as in Figure \ref{fig1}, but the
  background is now $\Lambda$CDM with $\Omega_{\Lambda}=0.7$.  The
  lower panel shows the corresponding distance moduli.  The solid
  line in this plot again shows the background, which is now
  $\Lambda$CDM, and the double-dot-dashed line shows EdS.}
\label{fig1L}
\end{figure}

In Figure \ref{fig3} we show the effect of considering voids at
different distances from us.  We consider three sets of voids, one set
that is $25\%$ under-dense today, another that is $50\%$ under-dense,
and a third that is $75\%$ under-dense.  The upper plot in this figure
shows how the maximum depth of void varies with redshift, for each set
of voids.  As expected, as the centre of the void is moved to greater
redshifts, the minimum density experienced by the photon increases.
This is due to the depth of voids increasing with time.  All the voids
considered here are $0.1$ redshifts wide today. We do not consider voids that are centred at $z<0.1$, as we
are interested in the effect of distant voids, and not the effect of us
living in a void, which is considerable in itself
\citep{inavoid1,inavoid2,inavoid3,inavoid4,inavoid5}.  In the lower plot of Figure
\ref{fig3} we show the maximum deviation in distance modulus, from the
background value, for each of the three sets of voids.  This maximum deviation can
again be seen to be proportionate to the maximum depth of void, as
experienced by the photon, and shown in the upper plot.  It can be
seen that the maximum deviation in distance modulus is highly
sensitive to the location of the void, and increases steeply at low $z$.

\begin{figure}
\begin{flushright}
\vspace{-15pt}
\epsfig{figure=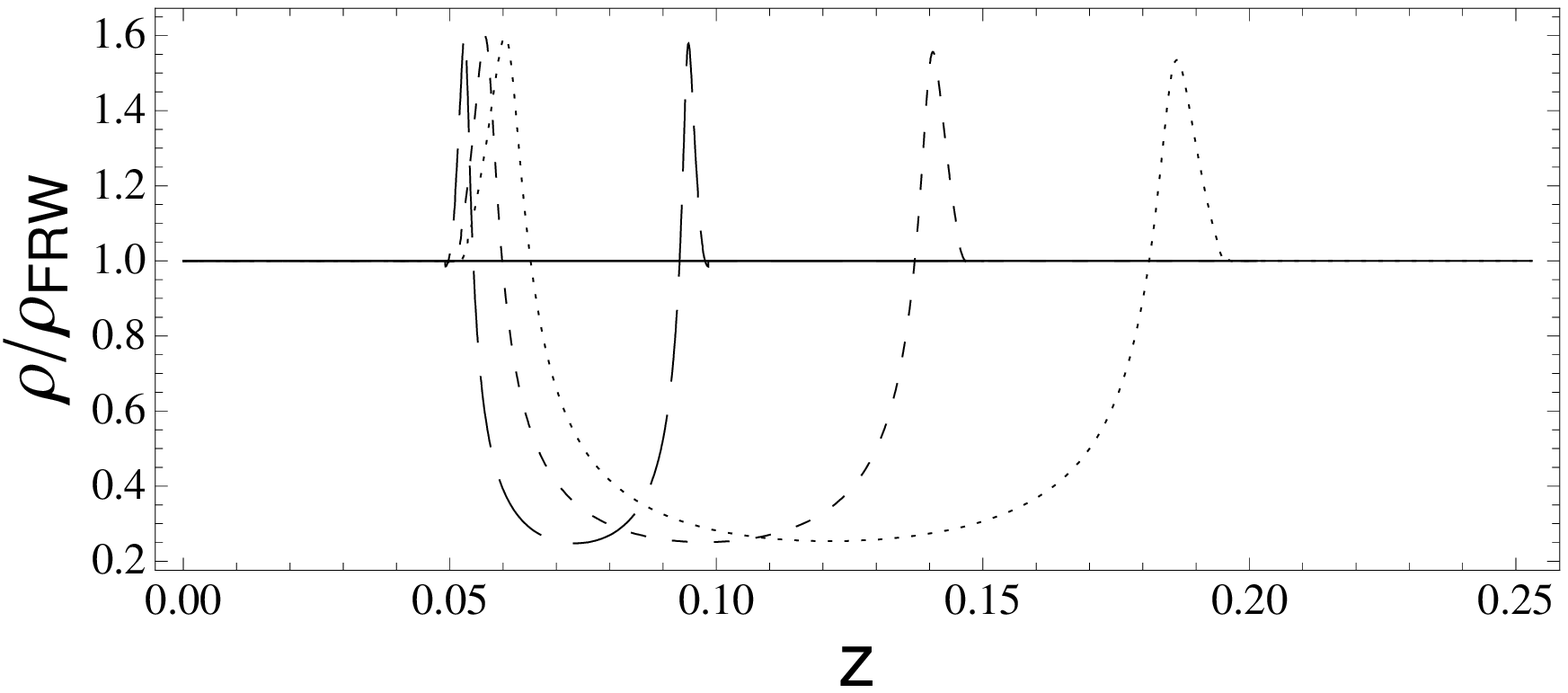,width=8.35cm}
\epsfig{figure=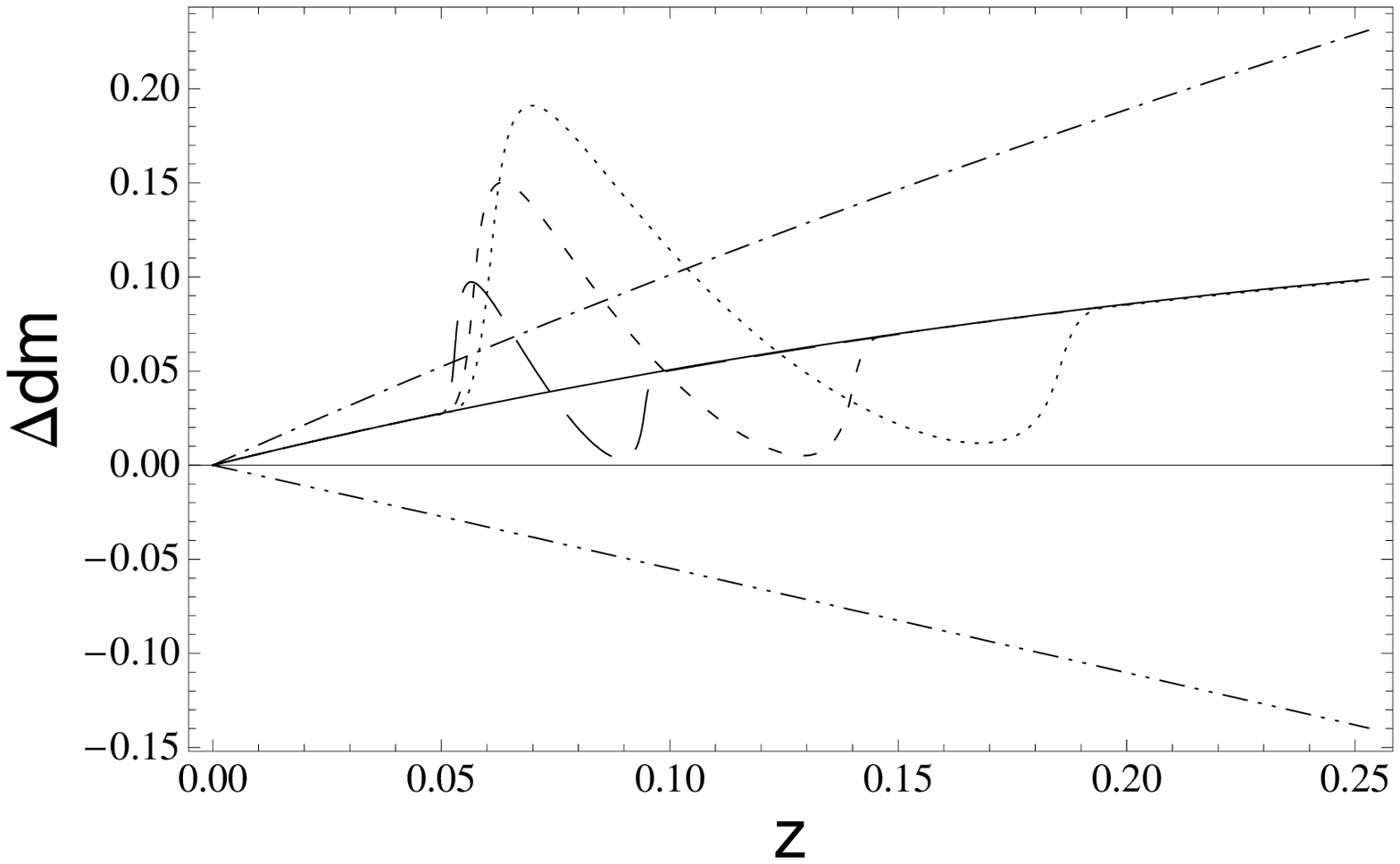,width=8.6cm}
\end{flushright}
\vspace{-10pt}
\caption{The upper panel is the same as Figure \ref{fig2}, but now with a
  $\Lambda$CDM background with $\Omega_{\Lambda}=0.7$.  The lower
  panel shows the corresponding distance moduli, with the solid and
  double-dot-dashed lines as in Figure \ref{fig1L}.}
\label{fig2L}
\end{figure}

\subsection{A $\Lambda$CDM background}

As well as a spatially flat, dust dominated EdS
background, we are also interested in backgrounds
containing a non-zero $\Lambda$.  To achieve an understanding of this we will repeat the
analysis above for the case of a single void in an
asymptotically $\Lambda$CDM universe with $\Omega_{\Lambda}=0.7$
\footnote{With radiation neglected, which should a good approximation in the epochs under
  consideration.}.  In this case the equations for $r(t)$, $z(t)$ and
$r_L$ are unchanged, with the functional form of $R(t,r)$ modified
from that given in (\ref{eds}) and (\ref{eds2}), in order to include
the effects of $\Lambda$ \citep{lambda}.

Figure \ref{fig1L} shows the same three voids as Figure \ref{fig1}, with under-densities of
$25\%$, $50\%$ and $75\%$ and centred at $z\sim 0.1$.  The
background is again given by a solid line, but this now corresponds to
$\Lambda$CDM with $\Omega_{\Lambda}=0.7$.  The EdS
distance modulus is shown by the double-dot-dashed line.  The
effect of the void can be seen to be similar that shown in
Figure \ref{fig1}, when $\Lambda=0$, but with a smaller magnitude.
The distance modulus returns to within $0.01$ magnitudes of the
background at large $z$, and there is a considerable, though smaller, displacement from
the background value when looking at objects inside the void.  It can be
seen that the energy density experienced by the photon is closer to
the value of the under-density today in this case, reflecting the fact
that the growth of structure slows when $\Lambda$ comes to dominate.

\begin{figure}
\begin{flushright}
\vspace{-15pt}
\epsfig{figure=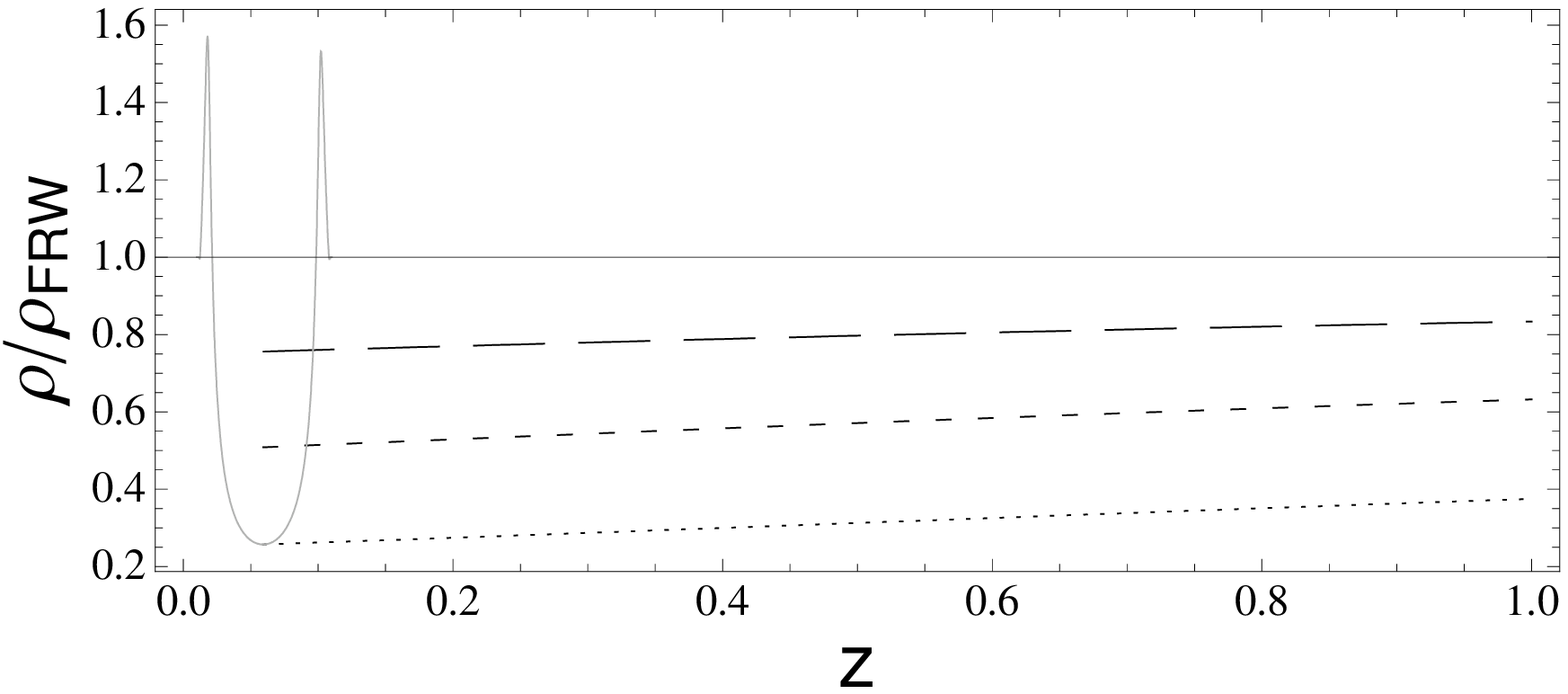,width=8.35cm}
\epsfig{figure=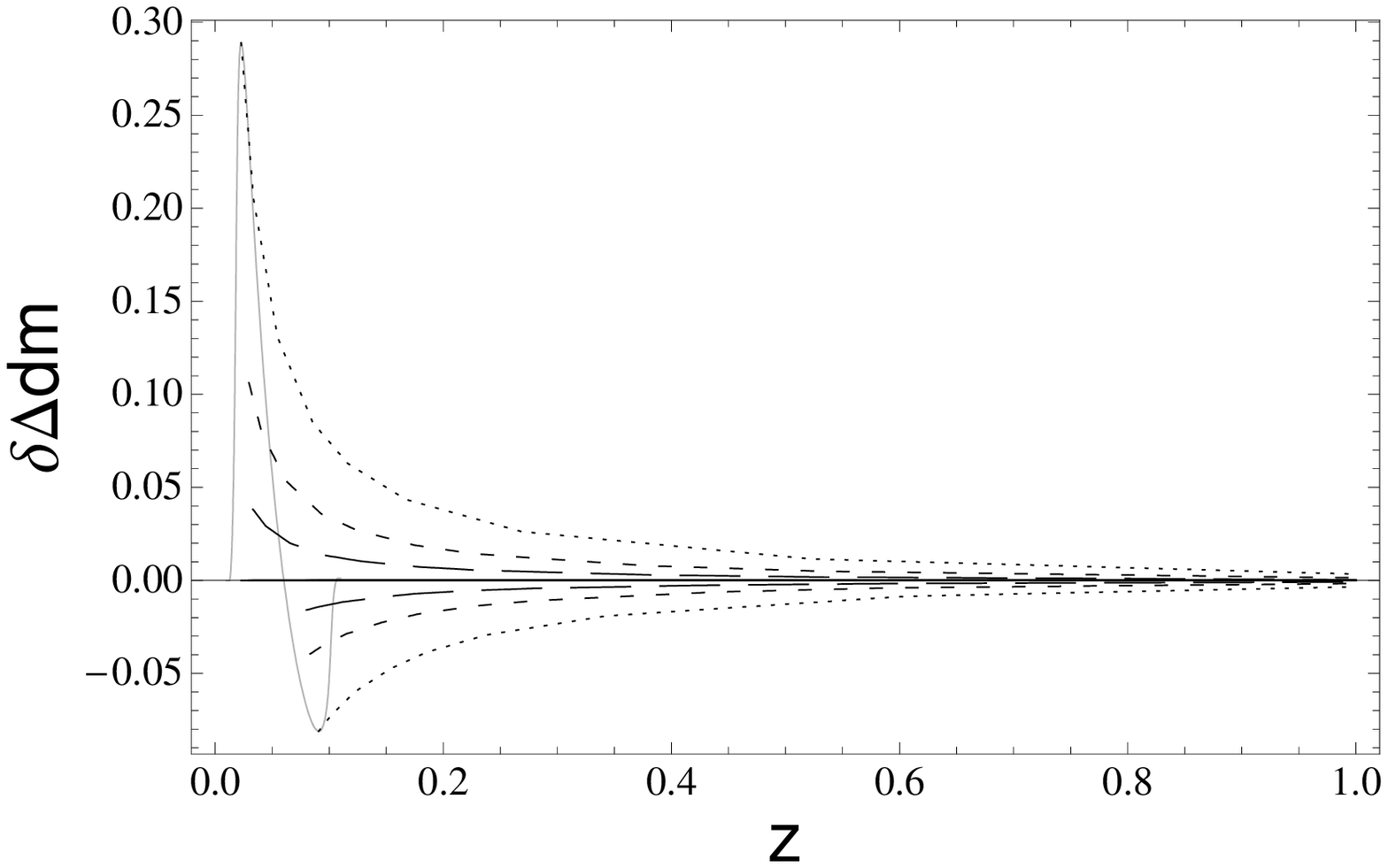,width=8.6cm}
\end{flushright}
\vspace{-10pt}
\caption{The same as Figure \ref{fig3}, but with a
  $\Lambda$CDM background with $\Omega_{\Lambda}=0.7$.}
\label{fig3L}
\end{figure}

Figure \ref{fig2L} illustrates the dependence of the distance modulus
on the width of void in this background, and the results can be seen
to be very similar to those deduced from Figure \ref{fig2}, but again
with a smaller overall magnitude of the displacement.  The
maximum deviation from the background is again sensitive to the
depth of the void, as well as its width.  In Figure \ref{fig3L} we
consider the maximum deviation of the distance modulus from its
background value, for the same three sets of voids as in Figure
\ref{fig3}.  The results are similar to the EdS case, the principle
differences appearing to be the smaller magnitude of the displacement,
and the weaker dependence of the depth of
under-density on $z$, due to the suppression of structure formation in
the presence of $\Lambda$.  This manifests itself in the upper plot of
Figure \ref{fig3L} as a slightly shallower gradient.

\subsection{Lensing}

The previous subsections have shown that at low redshift, in both EdS
and $\Lambda$CDM backgrounds, the principal effect of large voids is
on objects that are viewed within them.  In this case the displacement
of the distance modulus due to intervening voids is a smaller,
sub-dominant, effect.  At higher redshifts, however, the effect of
voids on objects within them is much smaller.  In this case the
lensing effect, due to looking through voids, becomes relatively more
important. In this subsection we quantify the size of this effect for
different sized voids, with different widths, at different redshifts,
in both EdS and $\Lambda$CDM backgrounds.

\begin{figure}
\begin{flushright}
\vspace{-25pt}
\epsfig{figure=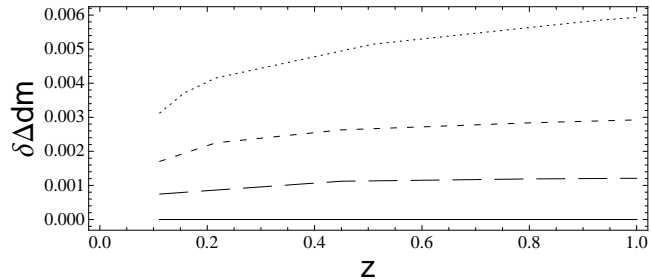,width=8.5cm}
\end{flushright}
\vspace{-30pt}
\caption{The displacement from the background distance modulus when
  looking through a single void, at different distances, in an EdS background.  The long-dashed
  line corresponds to voids with a central under-density of $25\%$
  today, the short-dashed line to $50\%$ today, and the dotted line to
  $75\%$.  The energy density experienced by the photon is given by
  the upper plot in Figure \ref{fig3}.  The values of $\delta
  \Delta$dm are plotted at the redshift that the photon leaves the void.
  All voids here have a width of $\Delta z = 0.1$.}
\label{NEWCDM}
\end{figure}

\begin{figure}
\begin{flushright}
\vspace{-25pt}
\epsfig{figure=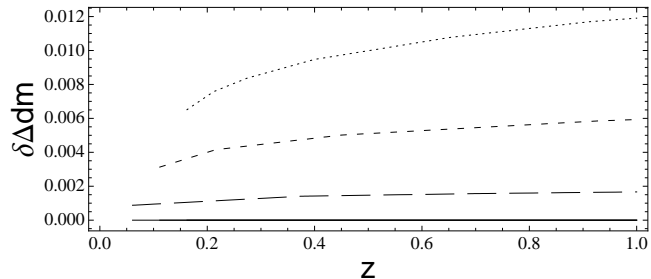,width=8.5cm}
\end{flushright}
\vspace{-30pt}
\caption{The same as in Figure \ref{NEWCDM}, but the voids considered
  are now all $75\%$ under-dense at their centre today.  The
  long-dashed, short-dashed and dotted lines now correspond to voids
  with widths of $\Delta z = 0.05$, $0.10$ and $0.15$.  These results
  are with as EdS background.}
\label{NEWCDMwidth}
\end{figure}

First let us consider an EdS background.  In Figure \ref{NEWCDM} we
plot the displacement from the distance modulus of the EdS background
due to a single void.  We consider three depths of void in this plot,
with central under-densities today of $25\%$, $50\%$ ad $75\%$.  All
the voids currently under consideration have a width, measured by an
observer standing at their edge today, of $\Delta z=0.1$.  The density
profile experienced by a photon for these three void depths, as they
are moved to different distances from the observer, are the same as
is shown in the upper plot of Figure \ref{fig3}.  What is plotted here
is the displacement from the background distance modulus at the moment
the photon leaves the void.
It is clear from Figure \ref{NEWCDM} that the lensing effect, while
smaller than the effects on the distance modulus shown in Figure
\ref{fig3}, is considerably less sensitive to the distance of the
void, and, in fact, increases somewhat as the void is moved further
away.  Increasing the depth of the void increases the magnitude of
this effect, in a similar way to was shown in Figure \ref{fig3}.

Figure \ref{NEWCDMwidth} shows the effect of looking through a single
void, at different distances, in an EdS background, but now varying
the width of the void, instead of its depth.  All voids are now $75\%$
under-dense at their centre today, but now have widths of $\Delta z =
0.05$, $0.10$ and $0.15$, depicted as the long-dashed, short-dashed
and dotted lines, respectively.  Again, $\delta \Delta$dm is given in
this plot at the moment the photon leave the void.  It can be seen that the
lensing affect is sensitive to the void width, and increases with the
width of the void.

\begin{figure}
\begin{flushright}
\vspace{-25pt}
\epsfig{figure=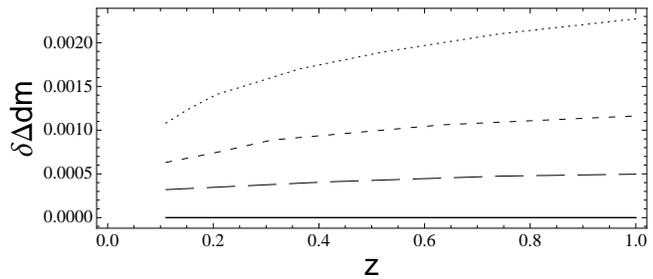,width=8.5cm}
\end{flushright}
\vspace{-30pt}
\caption{The same as Figure \ref{NEWCDM}, but in a $\Lambda$CDM
  background, with $\Omega_{\Lambda}=0.7$.  The long-dashed,
  short-dashed and dotted lines correspond to voids with $25\%$,
  $50\%$ and $75\%$ under-densities.}
\label{NEWLCDM}
\end{figure}

\begin{figure}
\begin{flushright}
\vspace{-25pt}
\epsfig{figure=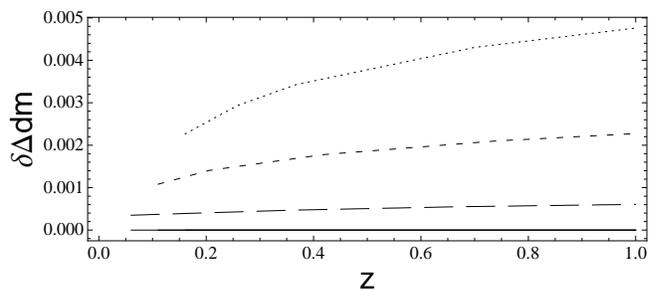,width=8.5cm}
\end{flushright}
\vspace{-30pt}
\caption{The same as Figure \ref{NEWCDMwidth}, but in a $\Lambda$CDM
  background, with $\Omega_{\Lambda}=0.7$.  The long-dashed,
  short-dashed and dotted lines correspond to voids with widths of
  $\Delta z=0.05$, $0.10$ and $0.15$.}
\label{NEWLCDMwidth}
\end{figure}

Now let us consider a $\Lambda$CDM background, with
$\Omega_{\Lambda}=0.7$.  The results of considering voids with
different depths, at different distances, are shown in Figure
\ref{NEWLCDM}.  The results are similar to the EdS case, but the
magnitude of the affect is smaller, in keeping with the results found
for looking at objects inside the voids.  In Figure \ref{NEWLCDMwidth}
we consider voids with different widths, and the same depth, at
different distances.  Again, the results are comparable to those found
in an EdS background, but with a smaller magnitude.

\section{Many Voids}
\label{manyvoids}

The effect of a single void on Hubble diagrams is of some interest by itself.
Such effects could be used to search for hypothesised large
structures, such as that which is supposed to explain the CMB cold
spot \citep{cold}. However, if we suppose that a single large void
exists in the Universe, then it is natural to think that there will be
other structure on similar scales.  In this section we consider just
such a scenario, with many LTB voids lined up back to back in an FRW background.

\subsection{An Einstein-de Sitter background}

Again, we will first consider the case of a spatially flat EdS
background.  Now instead of inserting a single void, satisfying
the boundary conditions (\ref{boundary1}) and (\ref{boundary2}), we
will insert a number of voids.  As before, we will match these voids
to the background cosmology at $r=r_0$, where they are locally FRW.  The
choice of voids that can be used for this process is quite arbitrary.  Here we will
randomly select voids from a couple of different distributions, in
order to gain some understanding.  This process can, of course, be repeated
for any distribution of voids one may wish to consider.

\subsubsection{Distribution A: Deep voids}

For this example we will consider voids with present day depths drawn from a
flat distribution between $0\%$ under-dense and $75\%$ under-dense
today.  The width of voids will be drawn from a flat probability
distribution, and will be between $0.01$ and $0.19$ redshifts wide
when viewed by an observer standing at their edge today.
Although these choices allow for the existence of large structures,
we do not consider them extreme (i.e. they are only a fraction
under-dense, not completely empty).

\begin{figure}
\begin{flushright}
\vspace{-15pt}
\epsfig{figure=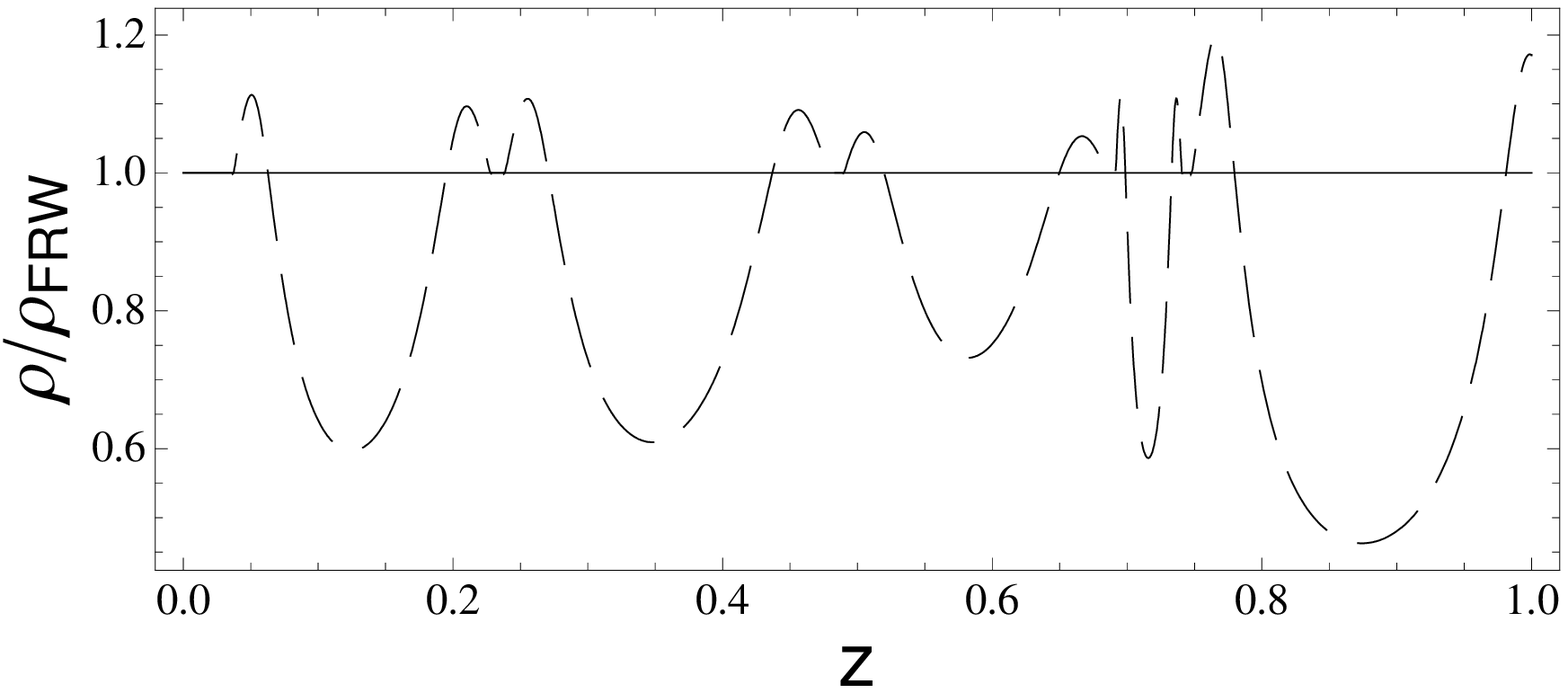,width=8.4cm}
\epsfig{figure=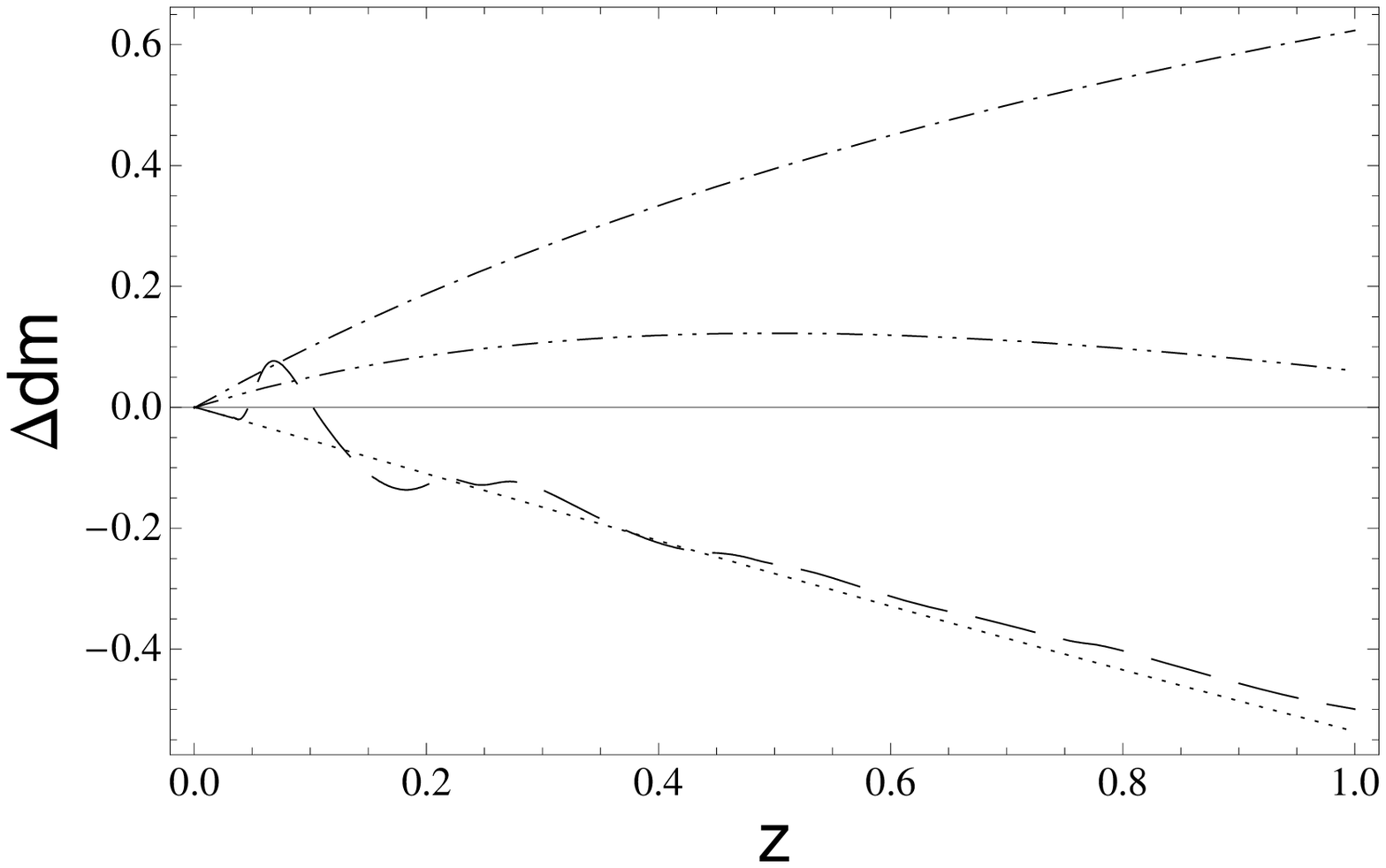,width=8.6cm}
\end{flushright}
\vspace{-10pt}
\caption{The upper panel shows a sample set of voids taken from the
  deep distribution of voids, A.  The lower plot shows the corresponding distance
  modulus.  The dot-dashed and double-dot-dashed lines are as in
  Figure \ref{fig1}. The dotted line is EdS.  These results are with an EdS background.}
\label{fig5}
\end{figure}

\begin{figure}
\begin{flushright}
\subfigure{\epsfig{figure=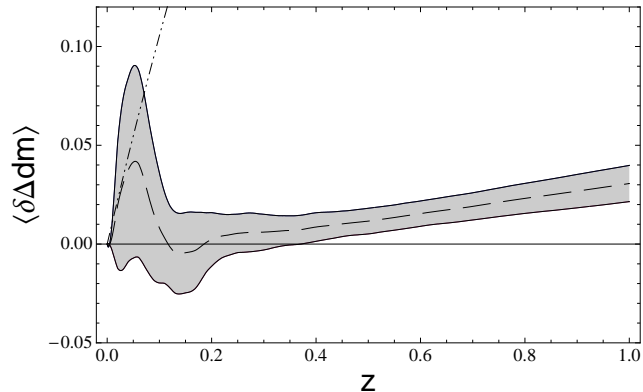,width=8.6cm}}
\end{flushright}
\vspace{-15pt}
\caption{The result of averaging the distance moduli
  of $1000$ sets of voids randomly drawn from the deep
  distribution, A.  Displayed is the mean deviation from the EdS
  background, the dashed line, and the standard deviation about that mean,
  the grey area.  Also displayed for reference is the deviation of
  $\Lambda$CDM with $\Omega_{\Lambda}=0.7$ from EdS, as the
  double-dot-dashed line.}
\label{fig4}
\end{figure}

In Figure \ref{fig5} we show an example set of voids, out to $z=1$,
picked from this distribution.  As was the case previously, the upper plot shows
the energy density encountered by a photon that reaches us at the
present day.  We ensure that we are not in a void by placing the edge
of the first void at a redshift of $z=0.01+w$, where $w$ is a random
number between $0$ and $0.05$ (the reason for introducing $w$ will be
made clear shortly).  The dotted line in the lower plot is
the background FRW universe, which here is EdS.  The dot-dashed lines
and double-dot-dashed lines represent the distance moduli of dS and
$\Lambda$CDM with$\Omega_{\Lambda}=0.7$, respectively.  While there
are clear deviations from the background distance modulus, it seems highly
unlikely that a distribution of voids of this type could be mistaken
for $\Lambda$CDM with $\Omega_{\Lambda}=0.7$.

\begin{figure}
\begin{flushright}
\vspace{-15pt}
\epsfig{figure=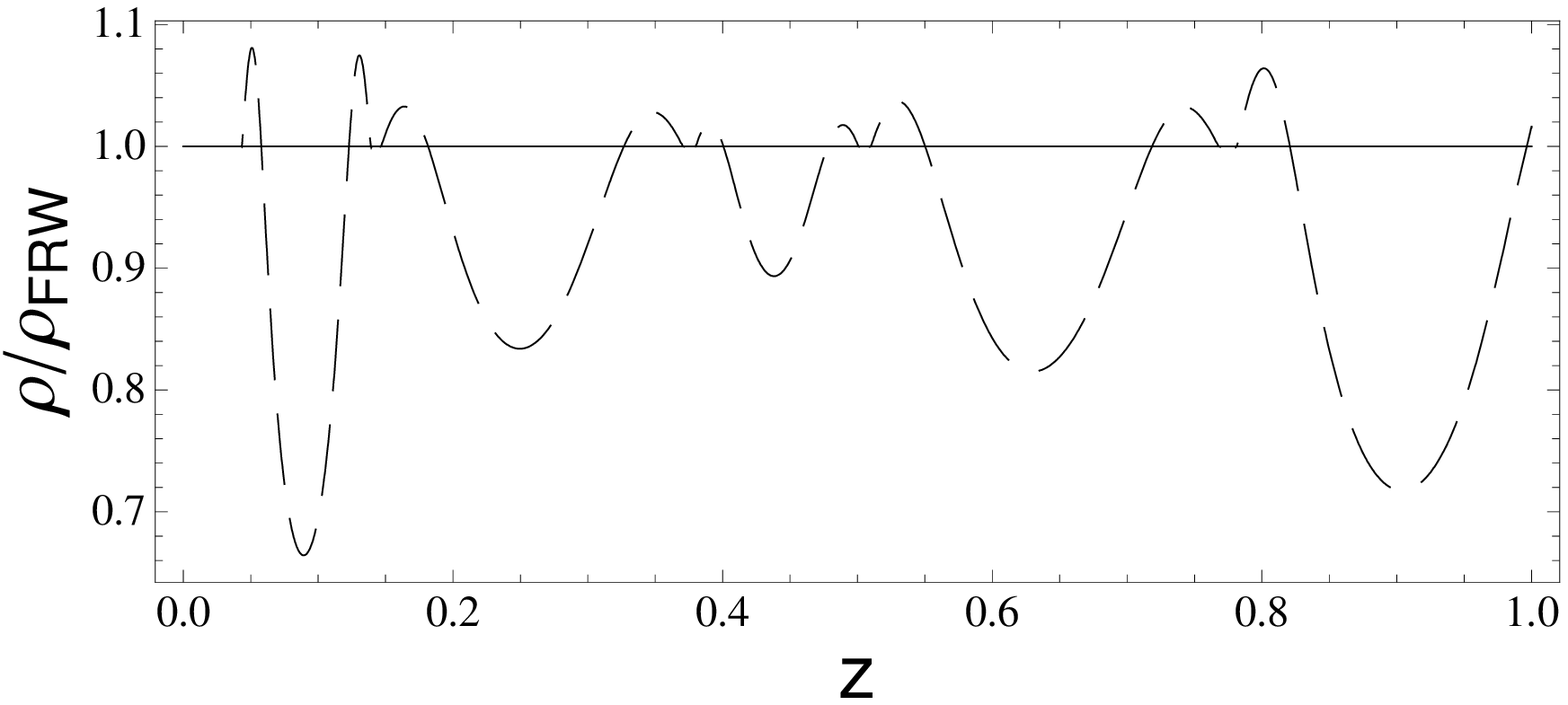,width=8.4cm}
\epsfig{figure=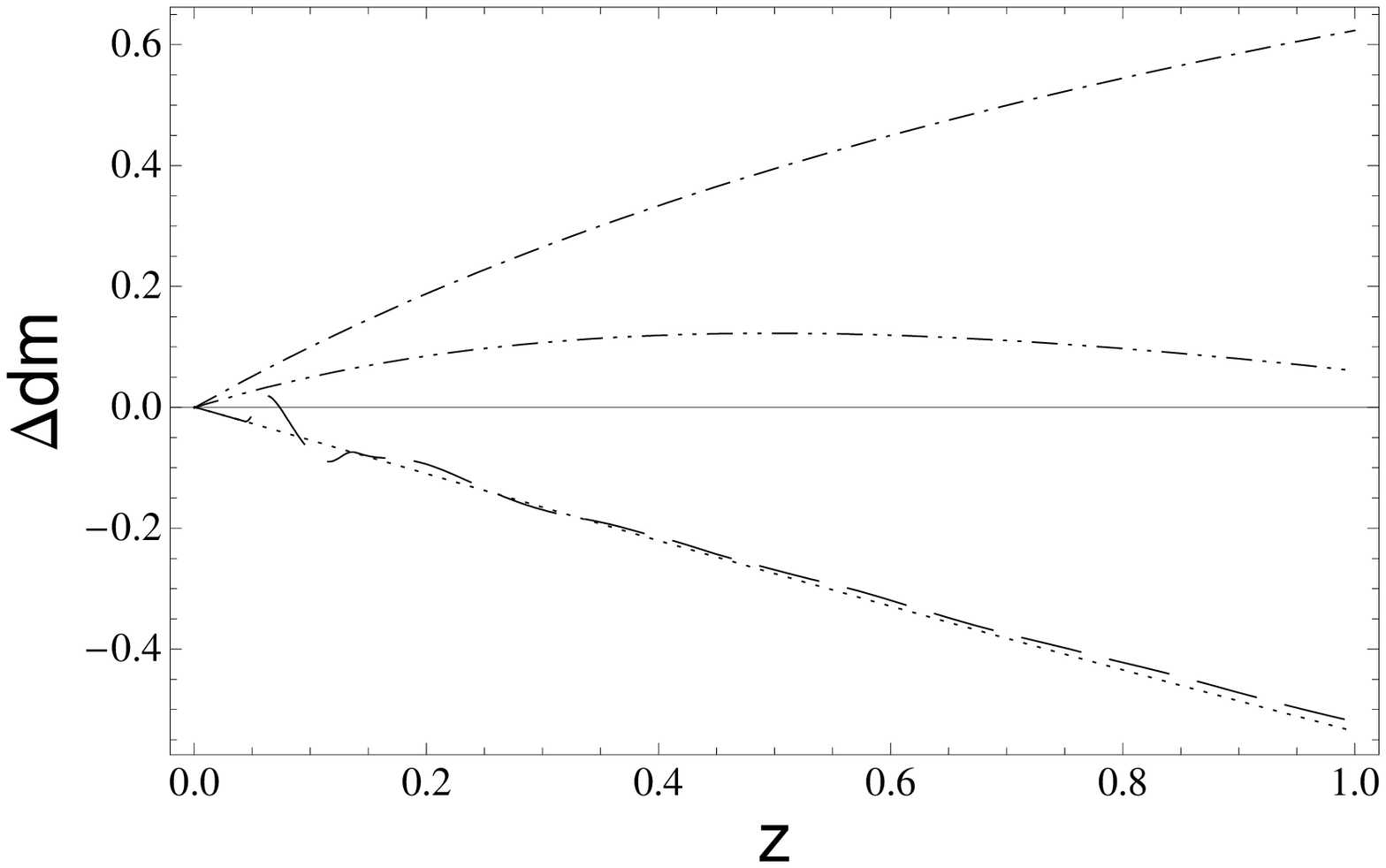,width=8.6cm}
\end{flushright}
\vspace{-10pt}
\caption{The same as Figure \ref{fig5}, but with voids drawn from
  the shallow distribution, B.}
\label{fig5b}
\end{figure}

\begin{figure}
\begin{flushright}
\subfigure{\epsfig{figure=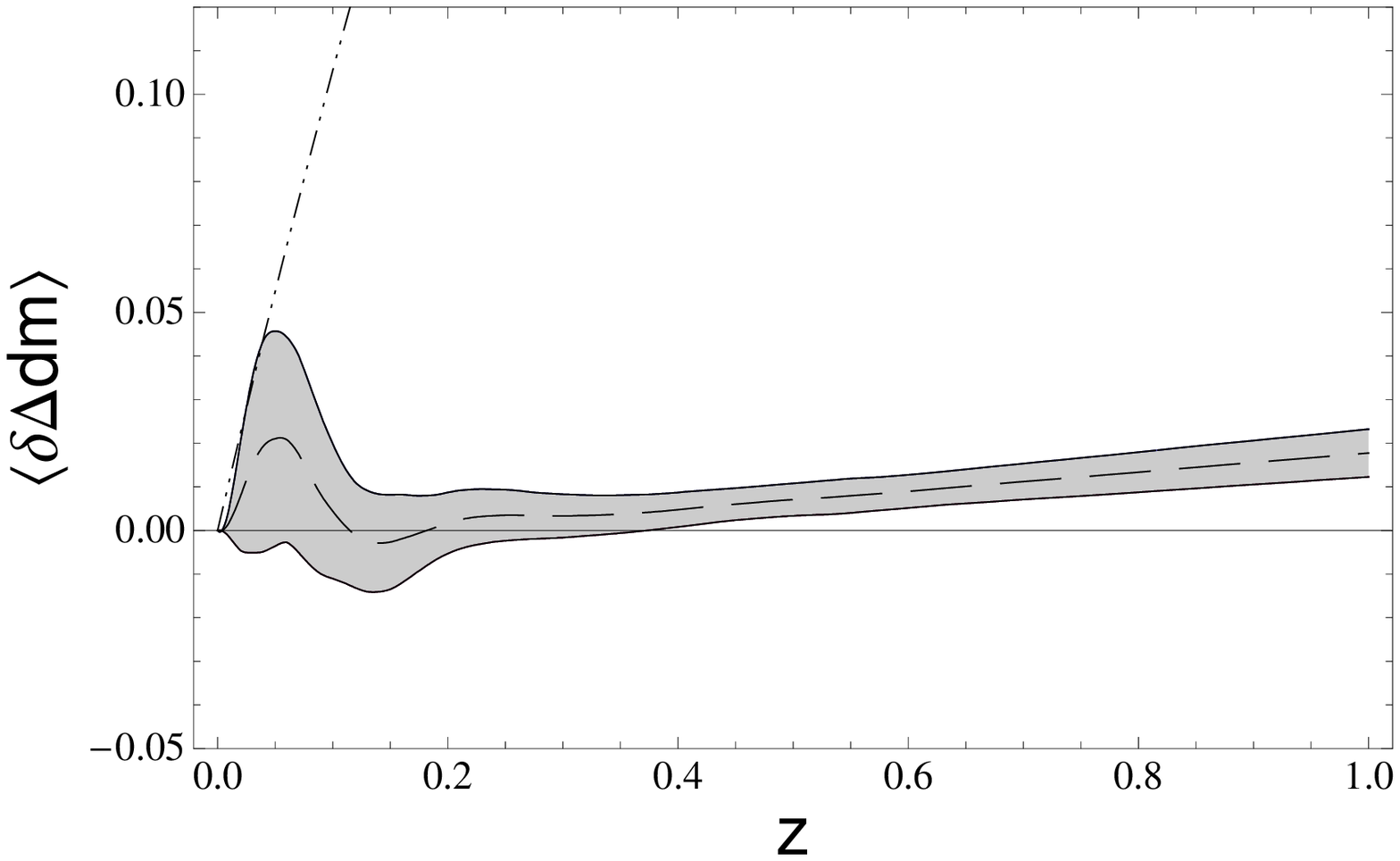,width=8.6cm}}
\end{flushright}
\vspace{-15pt}
\caption{The same as Figure \ref{fig4}, but with voids drawn from
  the shallow distribution, B.}
\label{fig4b}
\end{figure}

\begin{figure}
\begin{flushright}
\vspace{-15pt}
\epsfig{figure=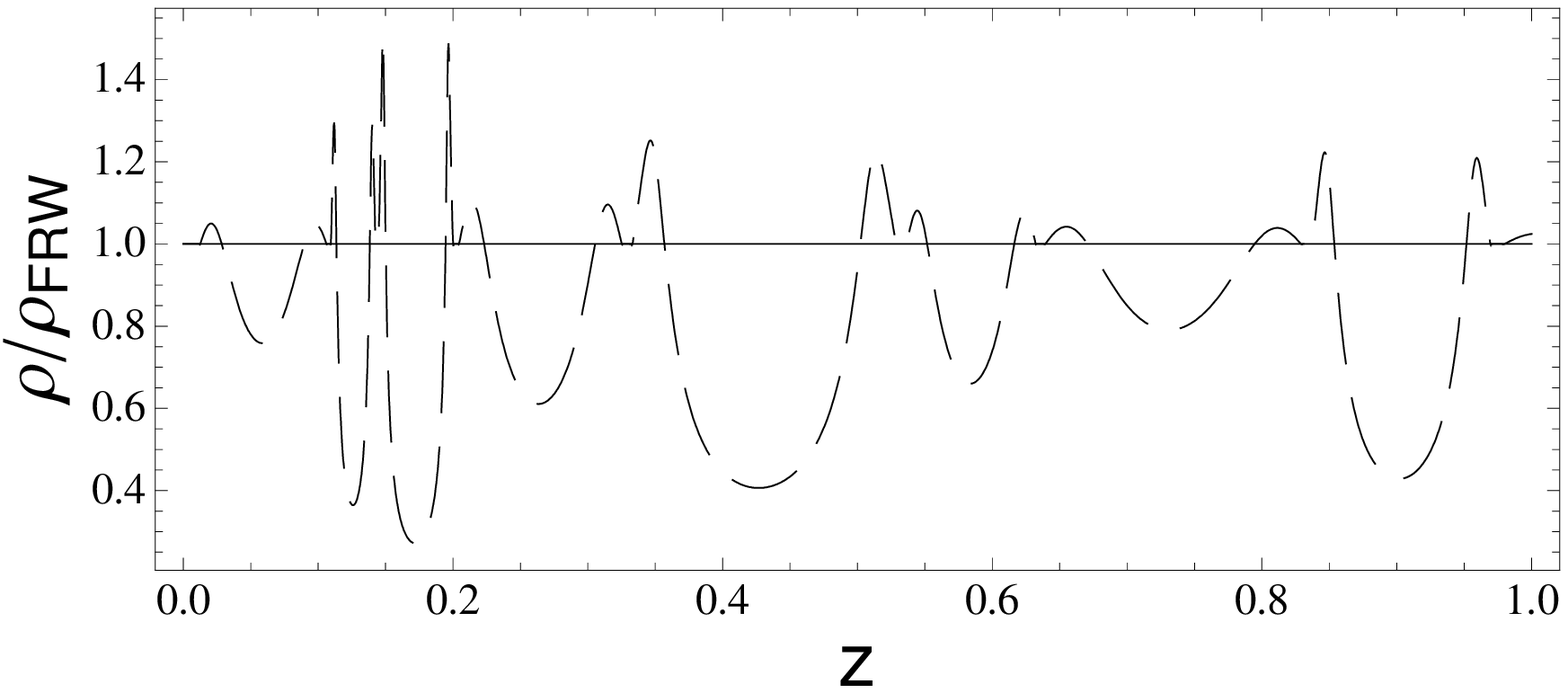,width=8.4cm}
\epsfig{figure=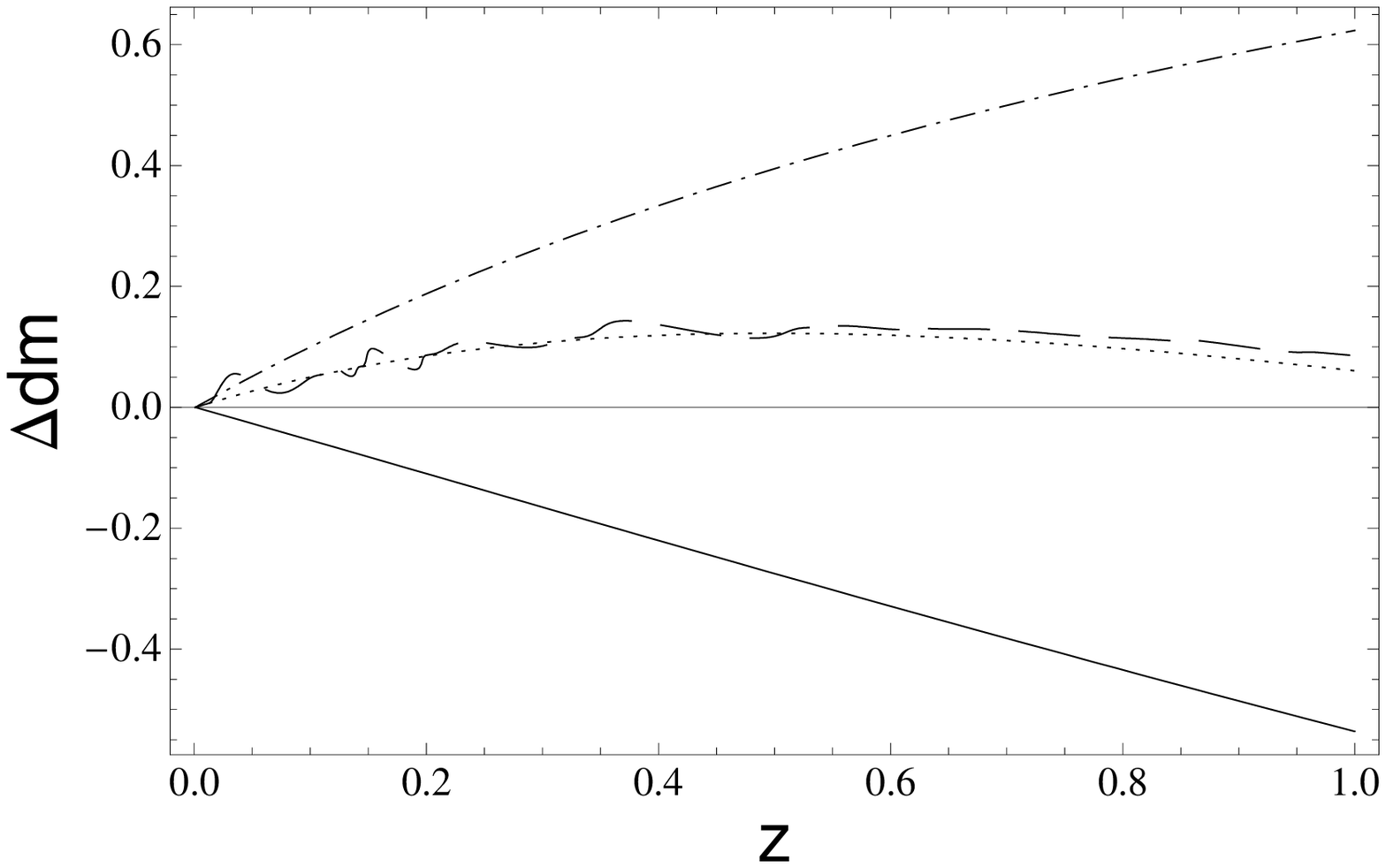,width=8.6cm}
\end{flushright}
\vspace{-10pt}
\caption{The same as Figure \ref{fig5}, but with a
  $\Lambda$CDM background with $\Omega_{\Lambda}=0.7$.  This set of
  voids is from distribution A.}
\label{fig5L}
\end{figure}
\begin{figure}
\begin{flushright}
\subfigure{\epsfig{figure=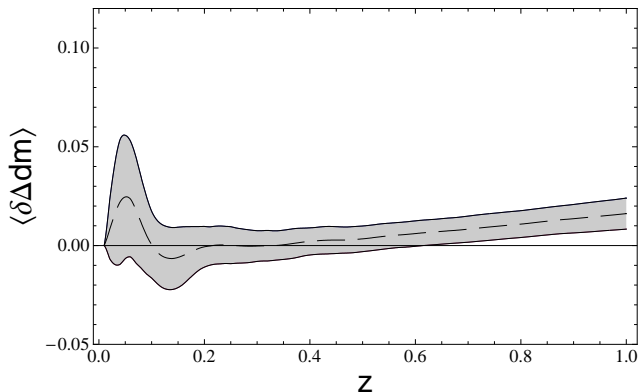,width=8.6cm}}
\end{flushright}
\vspace{-15pt}
\caption{The same as Figure \ref{fig4}, but with a
  $\Lambda$CDM background with $\Omega_{\Lambda}=0.7$.  This set of
  voids is from distribution A.}
\label{fig4L}
\end{figure}

Having considered an example set of voids, it is now of
interest to consider the average of many different sets drawn from the
same probability distribution.  This may be the type of process that
one would wish to consider when collecting observations made
over many different lines of sight, at different points on the sky.
Such an average is shown in Figure \ref{fig4}.  Here $1000$ different
sets of voids have been generated, their distance moduli calculated,
and an average taken.  Shown in the plot is the resulting mean
deviation from the background value, the central dashed line, and the
standard deviation from this value, the shaded region.  The effect
of $w$ here is to make the location of the edge of the first void
random (failing to include $w$ leads to a correlation in the location
of the first void, and an increased mean at that point).

The effect of looking at objects in an inhomogeneous universe of this
kind is two-fold.  Firstly, there is a systematic deviation
away from the distance modulus of the background.  This appears to
be largest at low redshift ($z<0.1$), when the effect of looking at
objects inside voids is considerable, and again at large distances
($z>0.5$), when the small lensing effect due to looking through voids
accumulates. These deviations are moderate for the void distributions chosen here, with 
$\left< \delta \Delta \textrm{dm} \right> < 0.1$, but are clearly
non-zero. Such an effect, if unaccounted for, could lead to a
systematic bias in extracting cosmological parameters.  Secondly,
there is a non-zero dispersion around the mean.  The effect of having a
distribution of voids between us and the source should be expected to
result in a typical source being somewhat displaced from the mean.
Again, the effect caused by this is largest at small redshifts ($z<0.2$).

\begin{figure}
\begin{flushright}
\vspace{-15pt}
\epsfig{figure=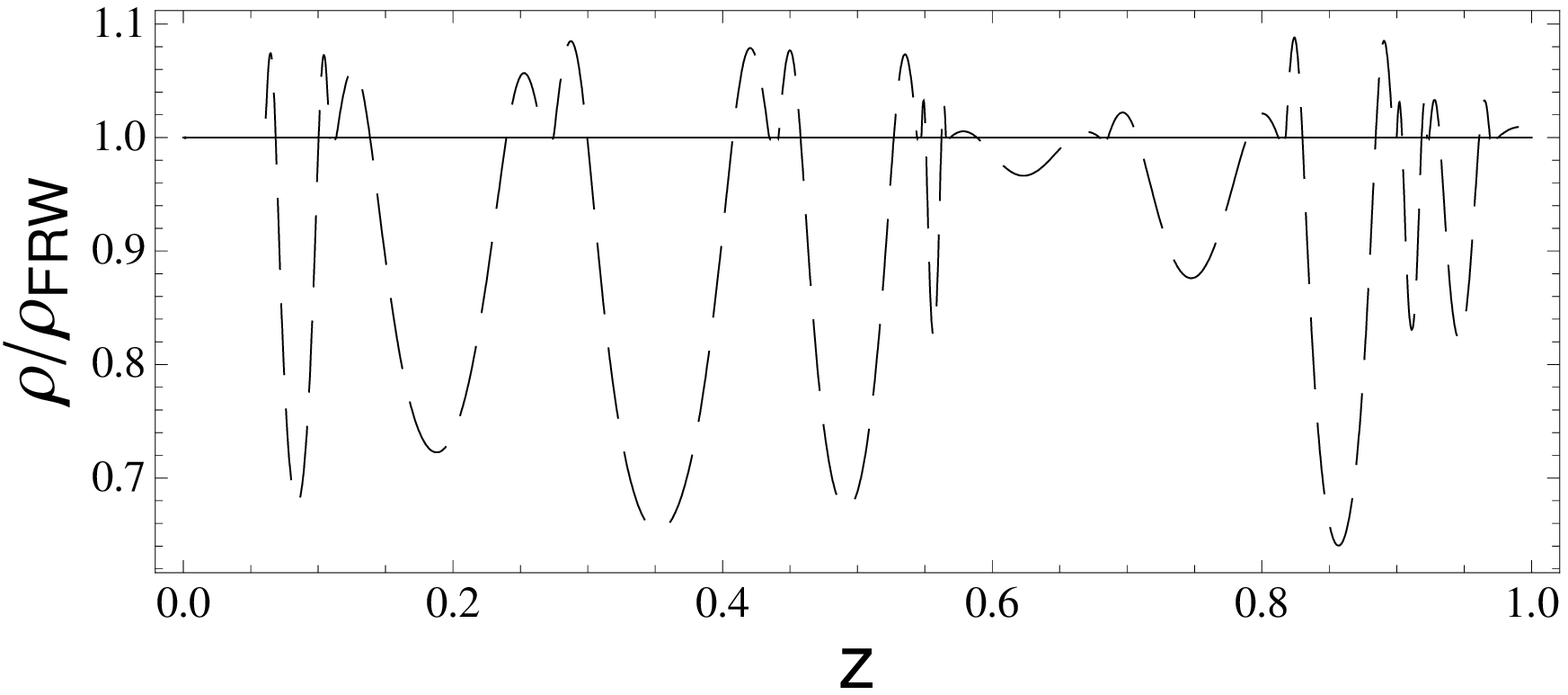,width=8.4cm}
\epsfig{figure=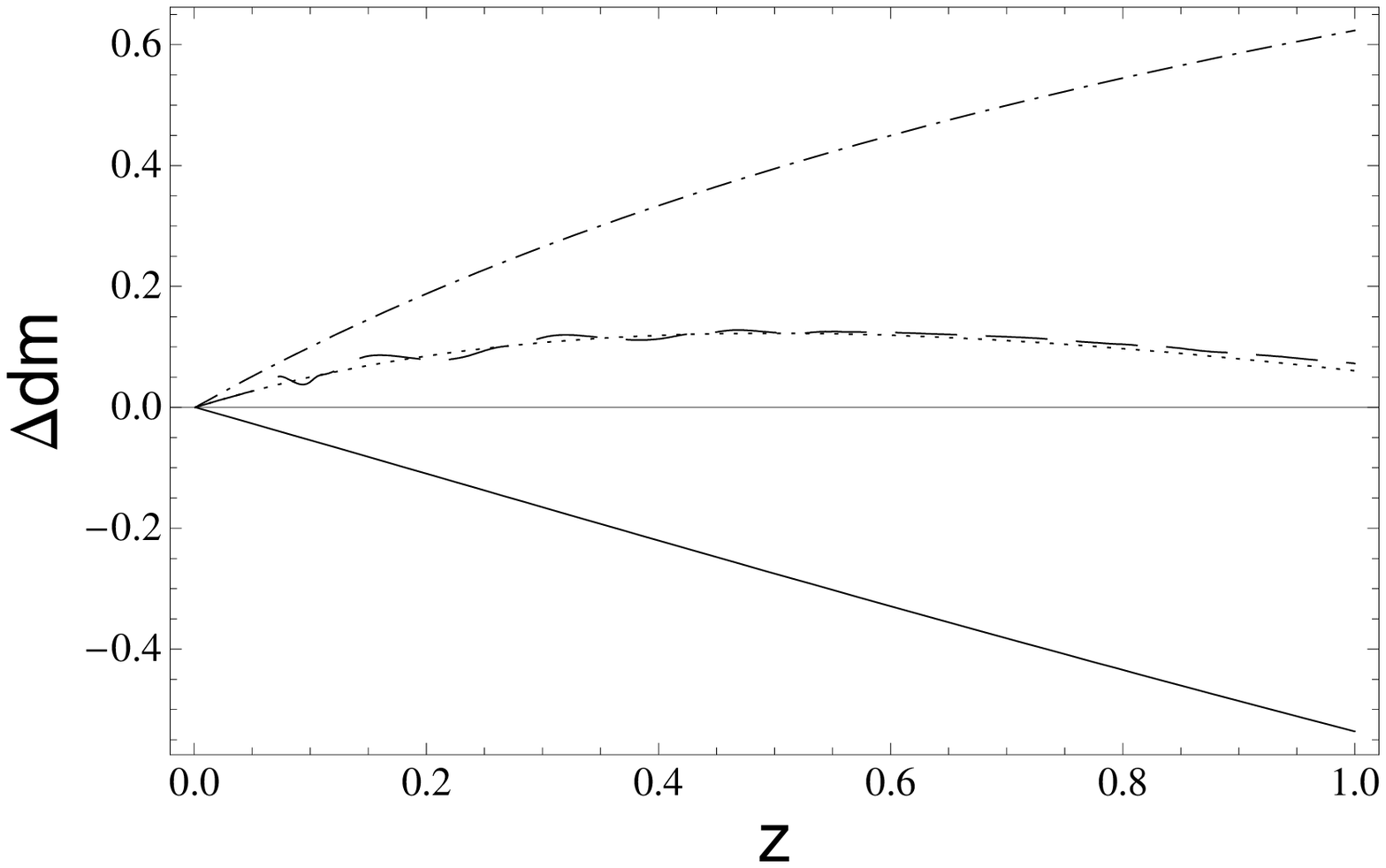,width=8.6cm}
\end{flushright}
\vspace{-10pt}
\caption{The same as Figure \ref{fig5L}, but with the
  shallow void distribution, B.}
\label{fig5Lb}
\end{figure}
\begin{figure}
\begin{flushright}
\subfigure{\epsfig{figure=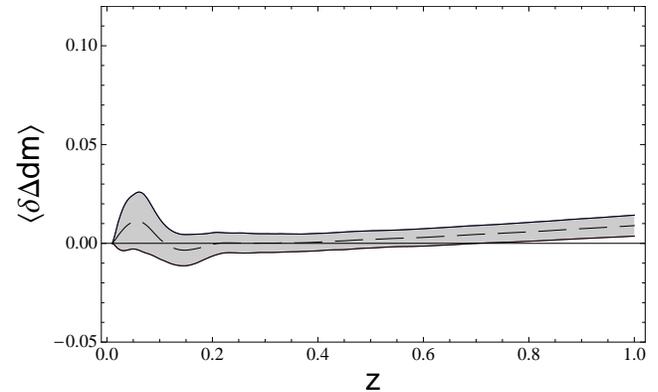,width=8.6cm}}
\end{flushright}
\vspace{-15pt}
\caption{The same as Figure \ref{fig4L}, but with the
  shallow void distribution, B.}
\label{fig4Lb}
\end{figure}

\subsubsection{Distribution B: Shallow voids}

Having considered one particular distribution of voids, let us now
consider a second so that we can better understand the effect of
the choice of voids on the resulting averaged distance moduli.  From the above
considerations of single voids, we know that the maximum displacement
of distance modulus is sensitive to the depth of void.  We will
therefore alter the probability distribution from which the void depth
is drawn.  We will now consider voids from a flat distribution that
are between $0\%$ and $50\%$ under-dense at their centre today, with
the same distribution of widths as before.
A sample density profile and distance modulus plot, for voids
drawn from this new distribution, is shown in Figure \ref{fig5b}.  This
can be seen to be similar to the results shown in Figure
\ref{fig5}, but with the deviations from the background value
reduced.

In order to find the mean, and standard deviation, for this new
distribution we will proceed as before.  $1000$ sets of voids
are generated, their distance moduli calculated, and then averaged.
The results of this are shown in Figure \ref{fig4b}.  These results
are again similar to those obtained from the previous
distribution, except with the magnitude of the deviations from the
background model decreasing in a proportionate way to the void depth.
The scatter around the mean is similarly decreased.

\begin{figure}
\begin{flushright}
\subfigure{\epsfig{figure=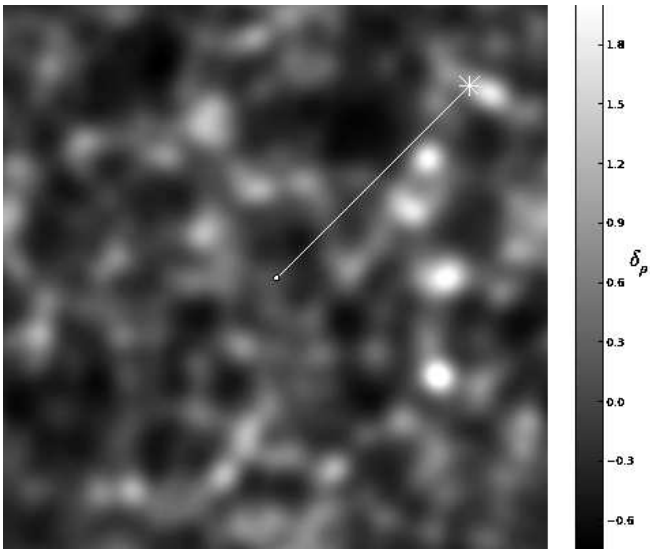,width=8.6cm}}
\end{flushright}
\vspace{-15pt}
\caption{A slice through the Millennium Simulation, smoothed on scales
  of 10Mpc.  The field is 715 Mpc across, the full size of the simulation. An example line of sight to a
  supernova is shown.} 
\label{fig-nonlinear}
\end{figure}

\subsection{A $\Lambda$CDM background}

Having discussed the situation of looking through many voids in an
EdS background, we will also be interested in space-times with a
non-zero $\Lambda$.  In this section we will consider a background
$\Lambda$CDM cosmology with $\Omega_{\Lambda}=0.7$. We will take voids
from the same two distributions considered above.

\subsubsection{Distribution A: Deep voids}

Using the same distribution of deep voids as above (that is with
central under-densities between $0\%$ and $75\%$ today), we
generate $1000$ sets of randomly selected voids.  An example set is
shown in Figure \ref{fig5L}.

In Figure \ref{fig4L} we show the result of averaging over $1000$
different sets of randomly generated voids.  What is shown here is the
deviation from the $\Omega_{\Lambda}=0.7$ background value.  As
before, the central dashed line shows mean deviation, and the shaded
region shows the standard deviation from this mean.  These results can be seen to be
similar to those shown in Figure \ref{fig4}, but with a smaller
displacement of the mean, and dispersion around that mean.  This is in
keeping with the result found in the single void case, that the
deviation of the distance modulus, due to the voids, is smaller with
non-zero $\Lambda$.

\subsubsection{Distribution B: Shallow voids}

Finally, let us consider $\Omega_{\Lambda}=0.7$ with the distribution of shallow voids, where
the central under-densities are between $0\%$ and $50\%$.  An example
set of voids, out to $z=1$, is shown in Figure \ref{fig5Lb}.  The
effect of these shallower voids is similar to the results found above,
but with smaller displacements of distance modulus from the background
value, as expected.

Figure \ref{fig4Lb} shows the results of averaging over $1000$ sets of
voids generated from this distribution.  As with the case of deeper
voids, the results are similar to those obtained when $\Lambda=0$,
shown in Figure \ref{fig4b}, but with smaller displacement of the
mean, and dispersion around that mean.

\begin{figure}
\begin{flushright}
\subfigure{\epsfig{figure=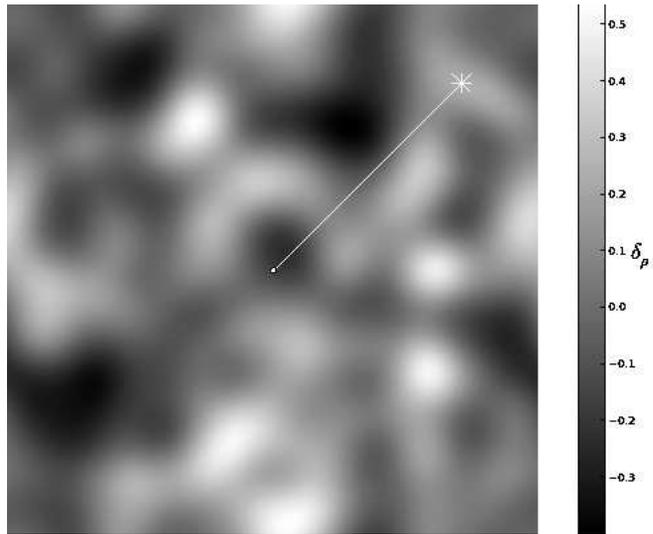,width=8.6cm}}
\end{flushright}
\vspace{-15pt}
\caption{The same Millennium Simulation slice as in Figure \ref{fig-nonlinear}, but smoothed on 20Mpc scales.}
\label{fig-nonlinear2}
\end{figure}

\section{Simulated Structure}
\label{realvoids}

In sections \ref{singlevoid} and \ref{manyvoids} we used exact Swiss Cheese cosmologies to
determine the effect of large inhomogeneities on luminosity
distances.  These situations are of interest as they allow
unambiguous, explicit calculations to be performed within them.
The draw-back of this formalism, however, is that while the theory is
non-perturbative, the mass distributions that can be modelled must be
highly symmetric.  This is in contrast to the linearised approach,
where the theory is approximated, but can then be applied more
straightforwardly to general mass distributions.  Despite these difficulties, we
will of course be interested in what the results we have found imply for more
realistic situations.

\begin{figure}
\begin{flushright}
\vspace{-15pt}
\epsfig{figure=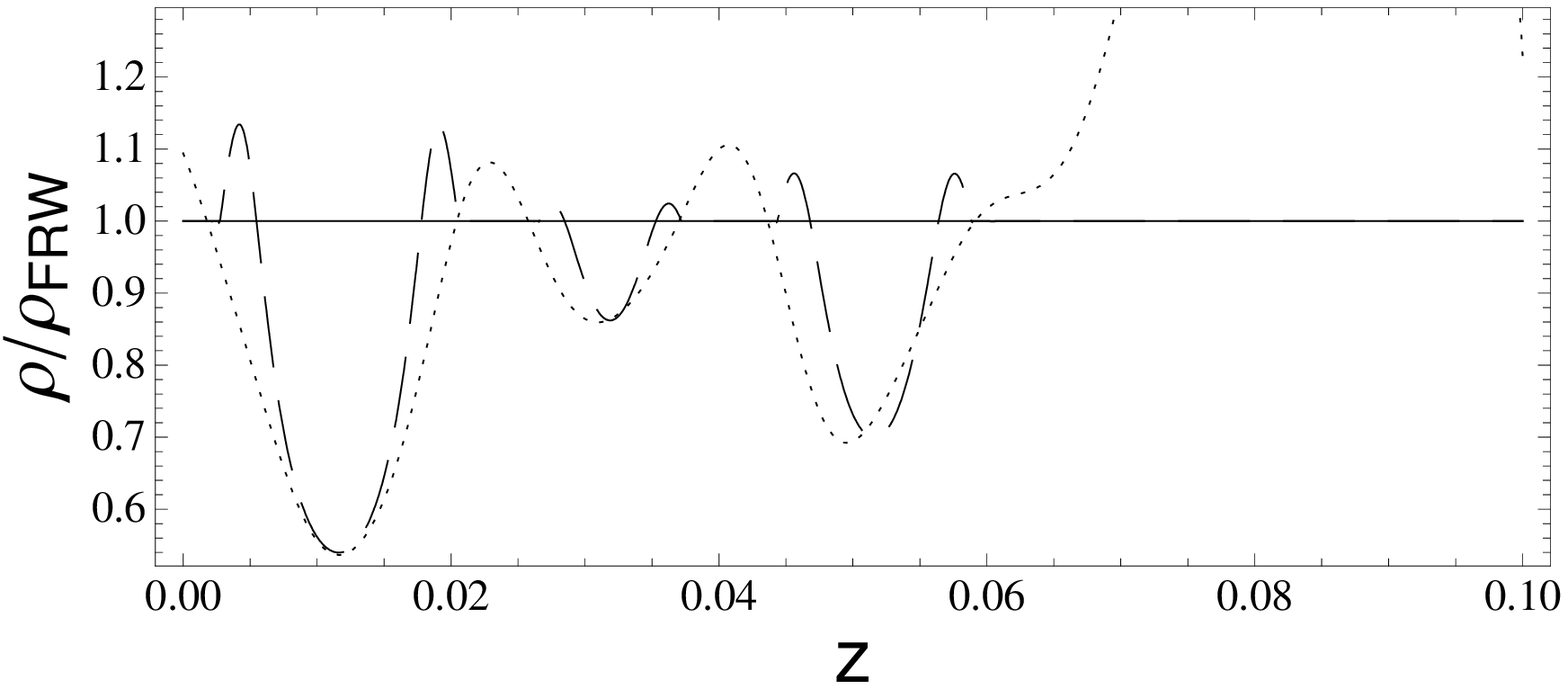,width=8.4cm}
\epsfig{figure=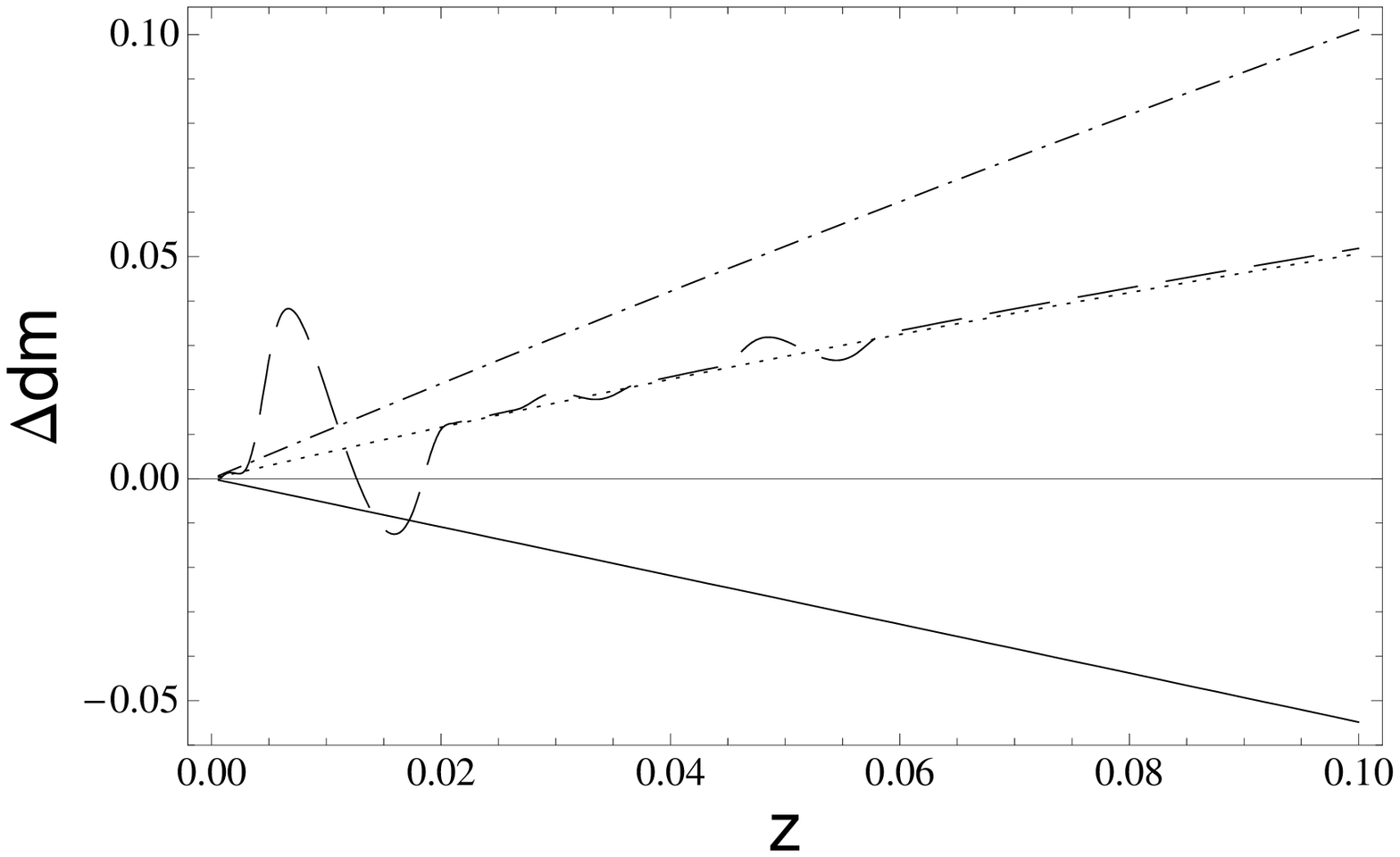,width=8.6cm}
\end{flushright}
\vspace{-10pt}
\caption{The upper panel shows the density along the line of sight
path shown in Figure \ref{fig-nonlinear} as the dotted line, and the
energy density of our idealised Swiss Cheese as the dashed line.
The lower panel shows the distance modulus generated by Swiss Cheese
as the dashed line.  The smoothing scale here is $10$Mpc.}
\label{triplot-nl-10}
\end{figure}

\begin{figure}
\begin{flushright}
\subfigure{\epsfig{figure=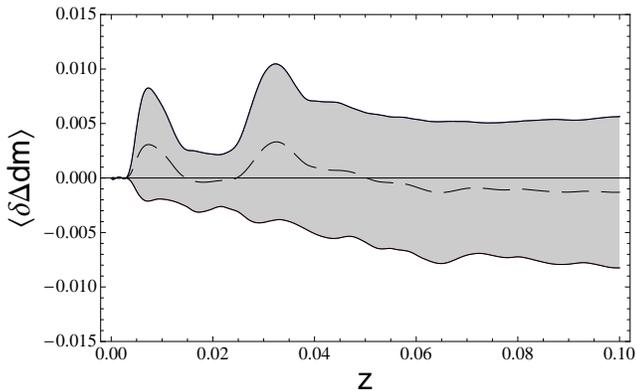,width=8.6cm}}
\end{flushright}
\vspace{-15pt}
\caption{The same as Figure \ref{fig4}, but for the Millennium
  Simulation smoothed on 10Mpc scales.}
\label{triplot-nl-10b}
\end{figure}

To address the problem of applying this formalism to more realistic
mass distributions we will take simulations of what real density fields in the Universe are
believed to look like.  These density fields will then be idealised so
as to appear as Swiss Cheese:  Over-dense regions will be taken as FRW
cheese, and under-dense regions will be modelled as LTB holes (with the appropriate width and depth).  The real
Universe, of course, is not an exact Swiss Cheese, and so
our idealisation will require some considerable approximation.  In
particular, under-densities will not in general be spherically
symmetric, but will have different expansion rates in different
directions.  Our goal here, then, must not be considered a precision
calculation of luminosity distances in these simulations, but rather
to derive a well motivated distribution of voids that can be used in
our Swiss Cheese models (rather than just considering idealised distributions).
One would hope that results derived in this way would be more
indicative of what may be expected in the real Universe.

\subsection{Density Fields}

\begin{figure}
\begin{flushright}
\vspace{-15pt}
\epsfig{figure=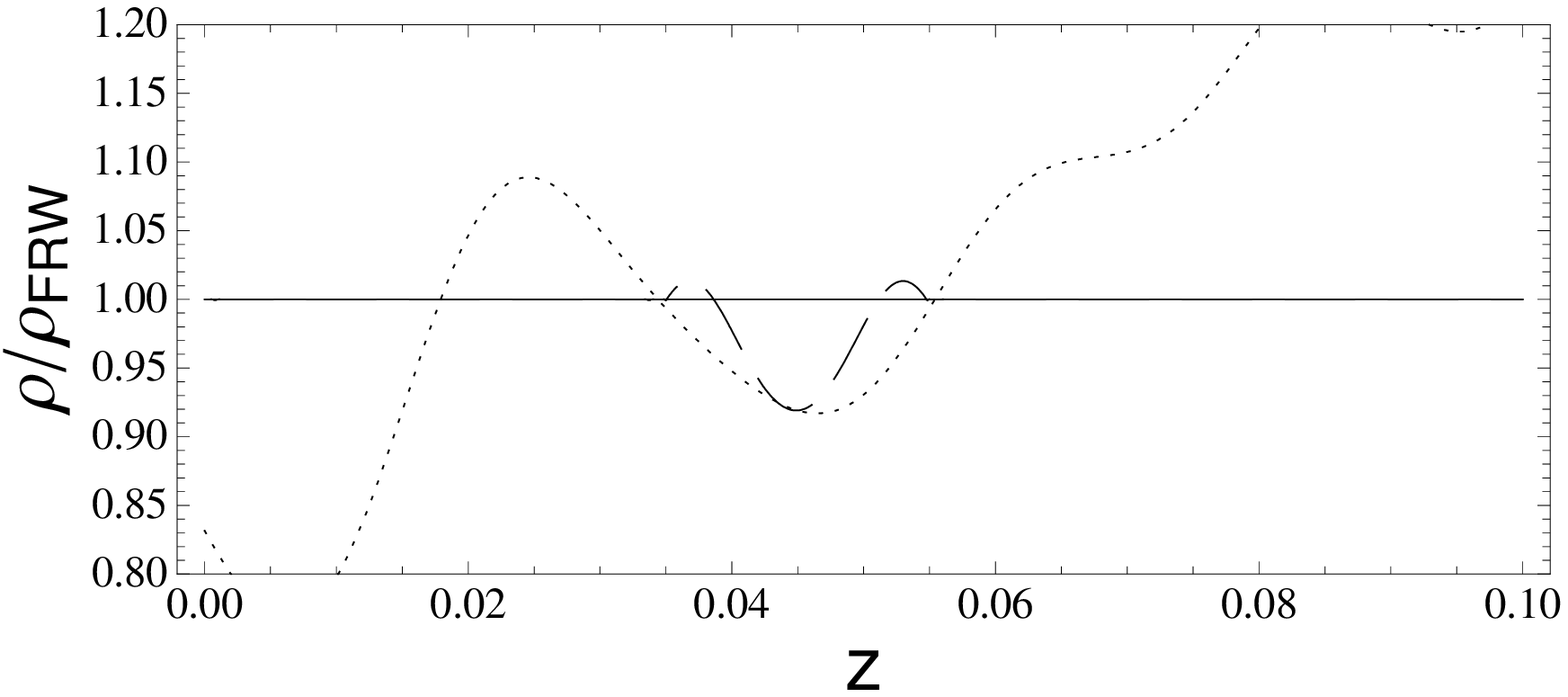,width=8.5cm}
\epsfig{figure=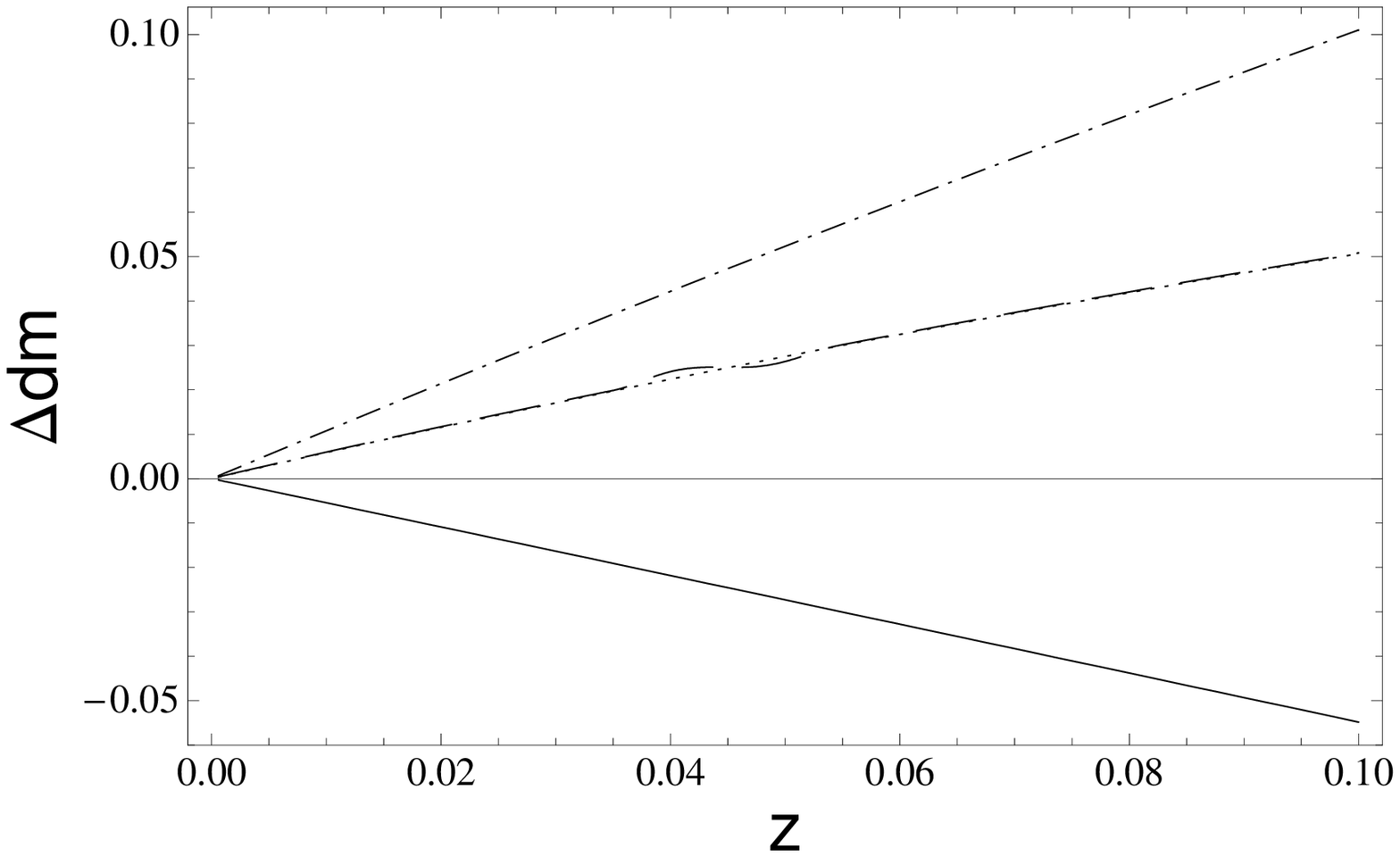,width=8.6cm}
\end{flushright}
\vspace{-10pt}
\caption{The same line-of-sight density profile, fitted voids, and
  distance modulus plots as in Figure \ref{triplot-nl-10}, but for
  the Millennium Simulation smoothed on 20Mpc scales.}
\label{triplot-nl-20}
\end{figure}

\begin{figure}
\begin{flushright}
\subfigure{\epsfig{figure=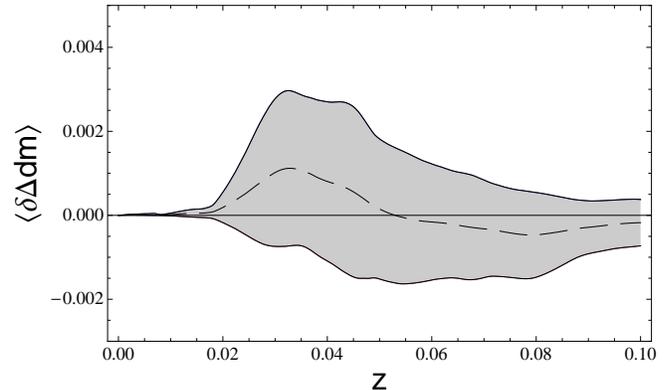,width=8.6cm}}
\end{flushright}
\vspace{-15pt}
\caption{The same as Figure \ref{triplot-nl-10b}, but for the Millennium
  Simulation smoothed on 20Mpc scales.}
\label{triplot-nl-20b}
\end{figure}


Extracting real density profiles from redshift surveys in the face of
limited survey regions, and redshift-space distortions, is somewhat
complex \citep{Lahav,erdogdu}.  In addition, the formalism we have used
up to this point takes density along a space-like projection of a
line of sight into a surface of constant $t$, whereas real
observations are made along past light cones.  We will therefore use simulations of
structure formation to generate line-of-sight
profiles of the energy density.  These simulations all have
$\Omega_{\Lambda}=0.7$ in their backgrounds.

While it is the case that the simulations we will be using are derived
within the regime of linear perturbations about an FRW background,
this does not stop us using these density fields to motivate
`realistic' distributions of voids.  The linear nature of the
underlying simulations suggests that, in fact, a linear treatment of
luminosity distances and redshifts should be adequate to accurately
calculate Hubble diagrams in this space-time.  The purpose of this
section is not to supersede such studies, but rather to
compliment them with a study of Swiss cheese models.

%


As a simulation of non-linear structure we use the results of the
Millennium N-body Simulation \citep{millennium1,millennium2}.  The
Millennium Simulation modelled a cubic region of space with a side
length of $500/h\,\, \textrm{Mpc}$, where $h\equiv H_0 / 100
\,\,\textrm{km s}^{-1} \,\textrm{Mpc}^{-1} = 0.7$.  It modelled $10^{10}$
dark matter particles in $256^3$ cells with periodic boundary
conditions, which for our purposes are especially suitable for
smoothing on new scales.  The database for the
simulation\footnote{http://www.mpa-garching.mpg.de/millennium/}
directly stores densities smoothed on various scales
from $1.25\textrm{Mpc}$ to $10.0\textrm{Mpc}$.



The size of density fluctuations is determined, at least in part, by
the smoothing scale applied to the data (the radius of the Gaussian kernel convolved with
the data).  Our choice of smoothing scale for the non-linear simulation is critical.
On small smoothing scales fluctuations are very large, but in these
regimes the pressureless dust approximation of the LTB model could break down.
In addition, on very small scales discrete particles (i.e. galaxies) may become
resolvable, and the fluid approximation itself could then be in question.
However, if we choose the smoothing scale to be too large then we will
under-estimate the density
fluctuations, and hence the effect on the distance modulus.  We use
smoothing scales of $10$ and $20$ Mpc; the former is direct from the
Millennium database and
the latter is manually smoothed with a Gaussian filter.  Slices
through the smoothed Millennium Simulation are shown in Figures 
\ref{fig-nonlinear} and \ref{fig-nonlinear2}.  The former has a
smoothing scale of $10$Mpc, and the latter is the same density field
with a smoothing scale of $20$Mpc.


\subsection{Distance Moduli}

In the density fields under consideration, lines of sight are
taken in random directions from the centre of the simulation.  We then
convert the density profiles along these lines into sets of
voids that can be used as input for the formalism developed above.
We approximate any under-dense segments of the line as
being produced by a void generated from a smooth negative perturbation
(\ref{kr}) in the curvature, $k$, with the same depth and width as the segment.  If an
under-dense region is split into two by a local maximum that is less
than half the depth of its shallowest neighbouring minima, then such
an under-density is considered as two voids back to back.
Over-dense regions are simply replaced by FRW geometry, with the same density as the
background.  We also ignore the effect of the local void that,
coincidently, happens to lie at the centre of the Millennium
Simulation.  As mentioned above, local voids have a different effect
to the distant voids we are concerned with here.

Let us consider the density fields obtained from
smoothing the $10^{10}$ particles of the Millennium Simulation on
different scales.  We will begin by considering a smoothing scale of
$10$Mpc.  In this case, the density profile along the example line of
sight from Figure \ref{fig-nonlinear} is shown in Figure \ref{triplot-nl-10}.
The mean displacement in distance modulus, and standard deviation
about that mean, are shown in Figure \ref{triplot-nl-10b} for a sample
of $1000$ lines of sight.

As mentioned above, the results we obtain in this section are strongly
dependent on the chosen scale of smoothing.  Of course, in reality there is only one local
expansion rate of space at any given point, but without any knowledge
of the smoothing scale to which this corresponds
it seems most prudent for us to illustrate the effect of choosing
different scales.   To this end, we will now consider the results
obtained from smoothing on a scale of $20$Mpc.  In this case the density profile
along the example line of sight from Figure \ref{fig-nonlinear2} is shown in Figure \ref{triplot-nl-20},
together with the corresponding distance modulus plot.

The mean and standard deviation of the difference in distance modulus
from the background value, for $1000$ lines of sight, are shown in
Figure \ref{triplot-nl-20b}.  This plot can be seen to be
significantly different to Figure \ref{triplot-nl-10b}: Doubling the
smoothing scale has more than halved the magnitude of the
displacement and dispersion.  Such strong dependence on the smoothing
scale shows that understanding the spatial variation of local expansion rates
in an inhomogeneous universe could be of critical importance in
determining its Hubble diagram.

Studies of luminosity distances in perturbed FRW space-times, such as
those of \citet{holz} and \citet{weak2}, find results that appear to
be of a similar order of magnitude, although derived in a very
different frame-work.  For example, \citet{holz} find that the maximum
dimming of the distance modulus from the background value is
$\delta \Delta$dm$\sim 0.1$ at $z=0.5$, as well as considerable
brightening of some sources due to lensing.  The study of
\citet{weak2} finds that a displacement of $\delta \Delta$dm$\sim
0.1$ could be achieved at larger redshifts, $z \sim 1$, and suggests
the interesting possibility of peculiar velocities having a
considerable effect on low redshift supernovae, with $z < 0.1$.

\section{Discussion}
\label{conclusions}

We have considered here the effects of large structures on
Hubble diagrams in a non-perturbative way.  Using the spherically
symmetric LTB exact solution of
Einstein's equations we have constructed a Swiss Cheese model of the
Universe.  In this model the background space-time is taken to be a
spatially flat FRW universe, filled with dust and a cosmological
constant.  Spherically symmetric regions of this background cosmology
are then removed and replaced with the regions of negatively curved LTB space-time,
matched appropriately at the boundary.  The resulting cosmology has an
initially evenly distributed energy density.  The regions with
negative spatial curvature then expand more quickly than the
background, and large under-dense voids form.  These voids behave like
open FRW space-time at their centre and transition smoothly to
over-dense regions at their edge, before matching onto the background.
Such a universe is an exact inhomogeneous solution of Einstein's equations, with a
simple enough geometry to allow calculations of
luminosity distances to be performed within it.

By considering the cross-sectional area of a bundle of radial null geodesics, focused on an observer in
the background space-time, it is possible to obtain a simple analytic expression for the
angular diameter distances such an observer will infer for objects in, and
beyond, the void.  The luminosity distance then follows
straightforwardly, and we can calculate the Hubble diagram that our
observer will see along such a line of sight.  The expressions used
are limited to spherically symmetric voids, and to lines of sight that
look directly through their centres, but have the great benefit that
they are exact.  As such, all non-linear effects due to position
dependent spatial curvature, expansion rates, and energy densities are
automatically included.

We then consider the effect of a single large void on Hubble diagrams
constructed by an observer looking through it.  We find that
observations of objects beyond the void are not noticeably affected by
its presence: Deviations from the background value are less than
$0.012$ magnitudes, for the voids considered.  When viewing objects within the void, however, the
effect of the voids presence can be of some size.  We illustrate the
differences witnessed by our observer by considering the distance
modulus of astrophysical objects.  This is given by the magnitude of
a source, minus the magnitude it would have at the same redshift in an empty
Milne universe.  We consider voids with an even distribution of gravitational
mass, and a smooth under-density in spatial curvature.  In
this case the magnitude to an object in the void is altered by
the voids presence.  This effect is most pronounced at low redshifts
(with $z <0.2$), while at higher redshifts it drops off, and becomes comparable
to the small effect of looking through voids.
The maximum displacement from the background distance modulus appears
proportionate to the fractional under-density at the centre of the
void, and decreases with increasing $\Lambda$.  This
can be seen from Figures \ref{fig3} and \ref{fig3L}, where the
relation between void depth and distance modulus displacement can be
seen to mildly super-linear. We note that this effect appears to be
largely due to $O(0.1)$ perturbations in the metric functionals,
and so may be less apparent in a perturbative treatment.

Having studied the case of a single void, we then move on to consider
a universe containing many voids.  In this case we draw the void widths and depths
randomly from probability distributions.  We consider a
shallow distribution of voids, in which the void depths are
between $0\%$ and $50\%$ under-dense at their centre, and and a deeper distribution, in
which they are up to $75\%$ under-dense.  By generating $1000$ sets
of voids from each distribution, we produce a mean deviation from the
background distance modulus, and the standard deviation that should be
expected from that mean.  As found in the case of single voids, the
displacement of the distance modulus from the background value is
proportionate to the depth of voids considered.  The dispersion is
similarly dependent on typical void depth.

Using data from the Millennium Simulation, we then proceed to consider how
structures in more realistic universes could affect the Hubble
diagrams constructed by observers in them.  To achieve this we take a
number of different lines of sight from a central position in the
simulation.  The density profile along these lines are determined by smoothing on different scales.
We then treat under-densities as being formed from spherically symmetric,
smooth, negative perturbations in $k$, and calculate the distance moduli
along lines of sight that pass through them.  Over-densities are replaced by regions of FRW
space-time, with the background density and expansion rate.  

The effects of the voids in the Millennium Simulation are small, but
non-zero.  We find that such voids could have a noticeable effect on the Hubble diagrams constructed by observers in
the space-time, by producing deviations and dispersion in the distance
modulus.  The magnitude of this effect, however, is strongly dependent
on the depth of voids, which is itself a function of how the smoothing
from discrete sources, to continuous fluid is performed.  Increasing
the smoothing scale decreases the depth of void, and hence
decreases the effect on the Hubble diagram.  Ultimately, it appears
that one needs to know the appropriate scale on which to smooth, if
one wants to make reliable quantitative predictions.  We do not
attempt to solve this problem here.

We find the non-linear effects of inhomogeneity in the Universe can
both displace the average distance modulus from its
background value, and introduce dispersion around that average.  These effects, if
improperly accounted for in fitting data to FRW models, could lead to
systematic errors in extracting cosmological parameters, and an
under-estimation of the errors involved.  Such an effect may be of use
for accounting for some of the `intrinsic error' usually added to
supernovae data when fitting to cosmological models, and which is
often a large fraction of the total error.  We find that the the
magnitude of these effects is sensitive to the depths of voids
involved, their width, and their distance from us.

The study we have performed here is limited in a number of respects,
and rather than being exhaustive is instead intended to be a thorough
investigation of the simplest case.  One will certainly be interested
in the effects introduced by different void profiles.  It will also be
of much interest to determine the effects of looking through voids in an
off-centre way.  It was suggested by \citet{flan} that such observations can be considerably
different to observations directly through the centre.  Generalising
the present study to off-centre observations is not trivial, but is
certainly possible, and we will consider this elsewhere. The studies
of \citet{biswas} and \citet{strong, strong2} suggest that considering
off-centre trajectories could have important effects.  There are also
issues of bias that one may wish to take into account in more detailed
studies.  For instance, it may be the case that more supernovae occur
in denser regions of the Universe, or that supernovae in less dense
regions are easier to observe. 

This study could be further generalised by considering non-spherical voids, and by better accounting
for over-dense regions of the Universe.  The LTB voids considered here
are only solutions as long as pressure is negligible.  A similar study
including anything approaching a realistic over-density will therefore
require solutions with more general fluid content, which are
considerably more difficult to find.  It is feasible that the inclusion
of over-densities in a more satisfactory way could cancel some of the
displacement effect, but it is hard to see how it could counteract
the dispersion.  In Appendix \ref{onion} we consider the effect of
having spherically symmetric shells of under-density, rather than
spherical holes (an Onion universe, rather than Swiss Cheese).  In
this case the effects of the under-densities produce deviations in the
distance modulus with the same order of magnitude, but with the
opposite sign, to Swiss Cheese.  This shows that a full understanding
of the way that large scale structure effects Hubble diagrams is
likely to be a complicated function of the detailed geometry of the
inhomogeneous Universe.  A specialised study of the effect on Hubble
diagrams of living in an Onion universe has been performed by
\citet{onion}, and agrees with what we find in Appendix \ref{onion}.

Finally, one may speculate on the effects of large scale structure as
potentially mitigating the need for Dark Energy.  We find that the voids required to mistake an
Einstein-de Sitter background for $\Lambda$CDM with
$\Omega_{\Lambda}=0.7$ would have to be very deep.  Even the void distribution considered
in Figure \ref{fig4b} does not come close.  The voids involved
would likely have to be more than $75\%$ under-dense at their
centre.  Even if such voids did exist, their ability
to mimic the shape of the distance modulus of $\Lambda$CDM would rely
on some fortuitous correlations in their positions, so that the mean
$\Delta$dm peaks at around $z\sim 0.5$ and drops off at lower
redshifts.  Alternatively, one may consider extending the present
study so that the background Cheese is spatially curved, or so that we are at the centre of a
void.  The effects of a local void have been shown by a number of authors to be able to
mimic, at least to some degree, the presence of $\Lambda$ on the
Hubble diagram \citep{inavoid1,inavoid2,inavoid3,inavoid4,inavoid5},
but requires a very deep and wide structure.

\section*{Acknowledgements}

We are grateful to C. Clarkson, P. Ferreira and J. Silk for helpful
suggestions and discussion. TC acknowledges the support of Jesus
College, and JZ acknowledges that of the STFC.  We are also
grateful to have received support from the BIPAC, and helpful
suggestions from an anonymous referee.


\appendix

\section{An Onion Universe}
\label{onion}

\begin{figure}
\begin{flushright}
\vspace{-10pt}
\subfigure{\epsfig{figure=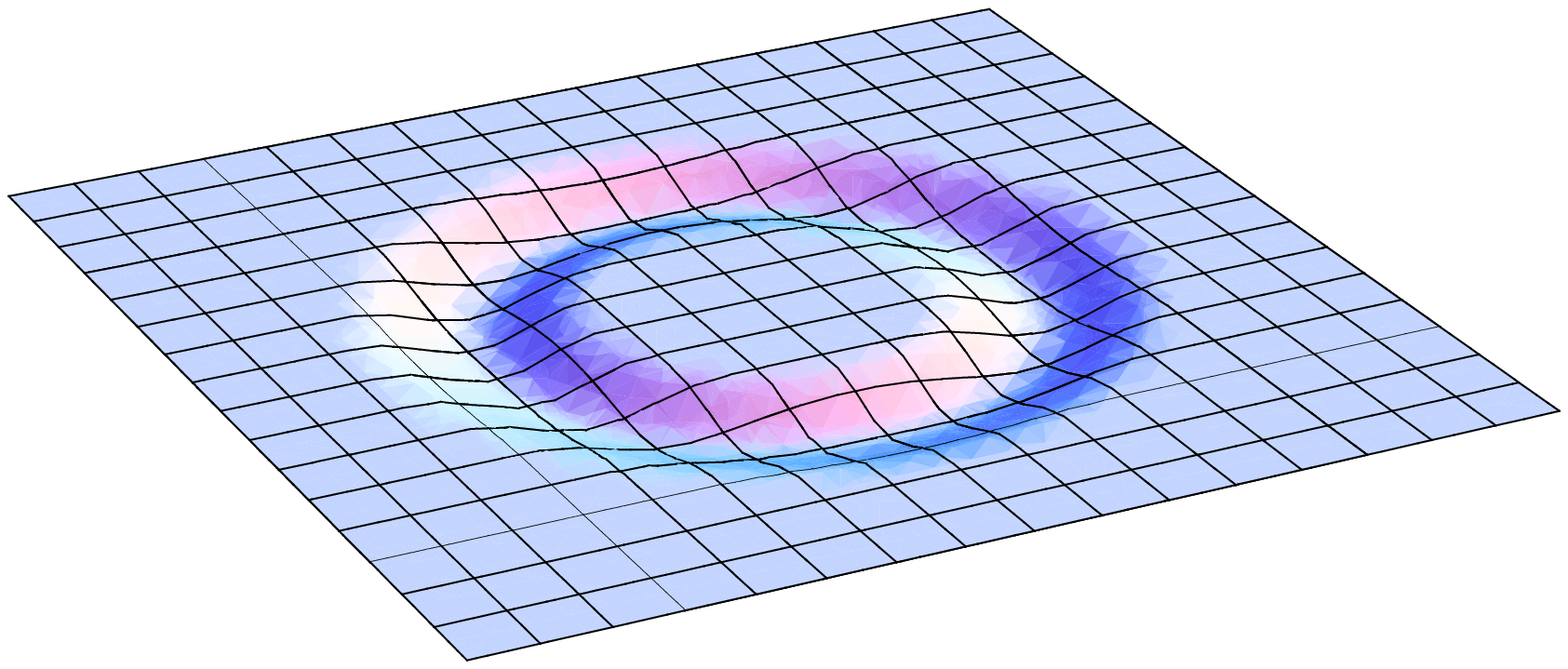,width=8.6cm}}
\end{flushright}
\vspace{-35pt}
\caption{An illustration of $k(\textrm{{\bf x}})$ in an Onion
  universe. The surface displayed is a constant time slice, with one
  spatial dimension suppressed.  Vertical displacement indicates a
  fluctuation in $k$, and distance from the centre is proportional to
  the coordinate distance, $r$.}
\label{toy}
\end{figure}

As an alternative to considering the Swiss Cheese model, one can also use the
LTB solution to consider an `Onion Universe', in which there are
under-dense shells of matter in a spherically symmetric space-time.
Although less appealing as a way of modelling the real inhomogeneous
Universe, a brief consideration of luminosity distances in such a
space-time will allow us to put the Swiss Cheese results in some
context.  An example of
the spatial curvature profile of an Onion universe is given in Figure
\ref{toy}, where a surface of constant time is shown, with one spatial
dimension suppressed.  Vertical displacement indicates a fluctuation
in $k$, and distance from the centre is proportional to comoving
distance, $r$.

\begin{figure}
\begin{flushright}
\vspace{-15pt}
\epsfig{figure=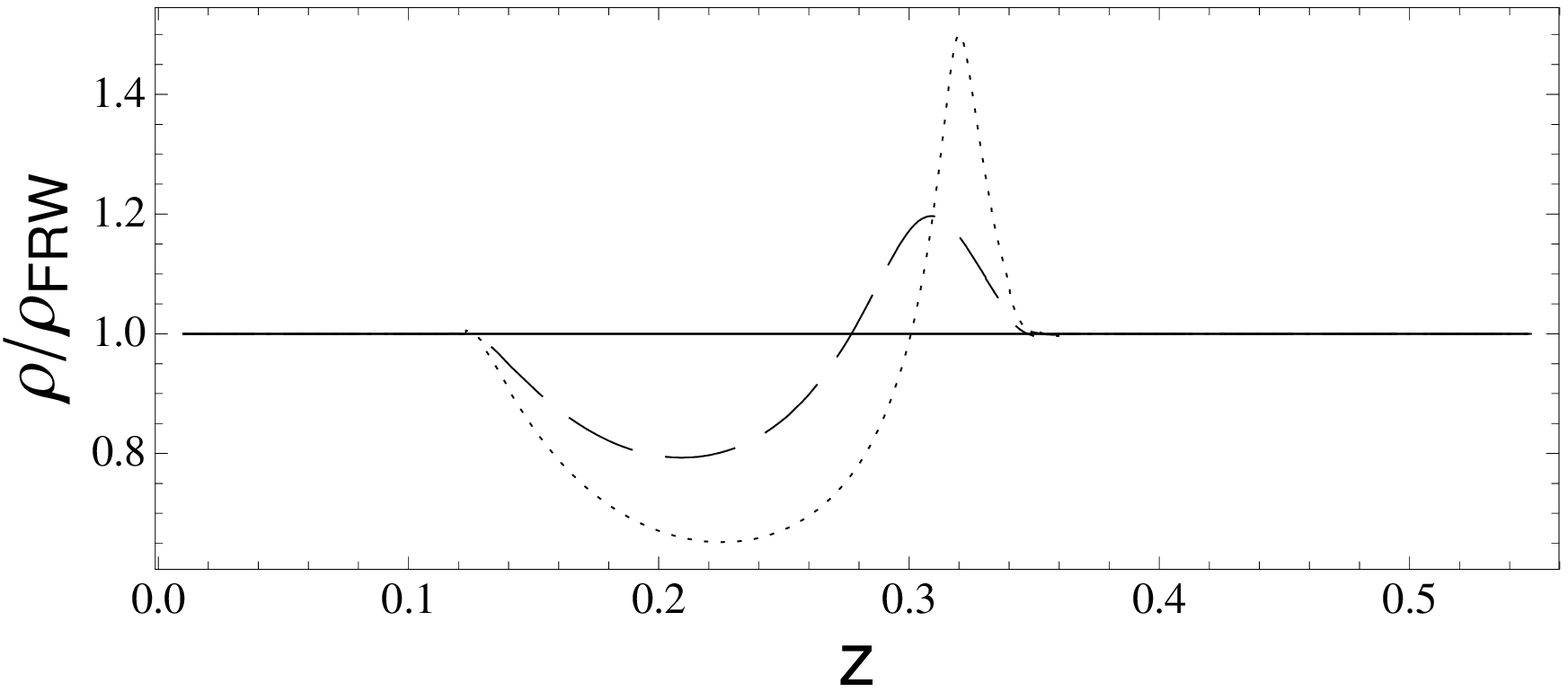,width=8.35cm}
\epsfig{figure=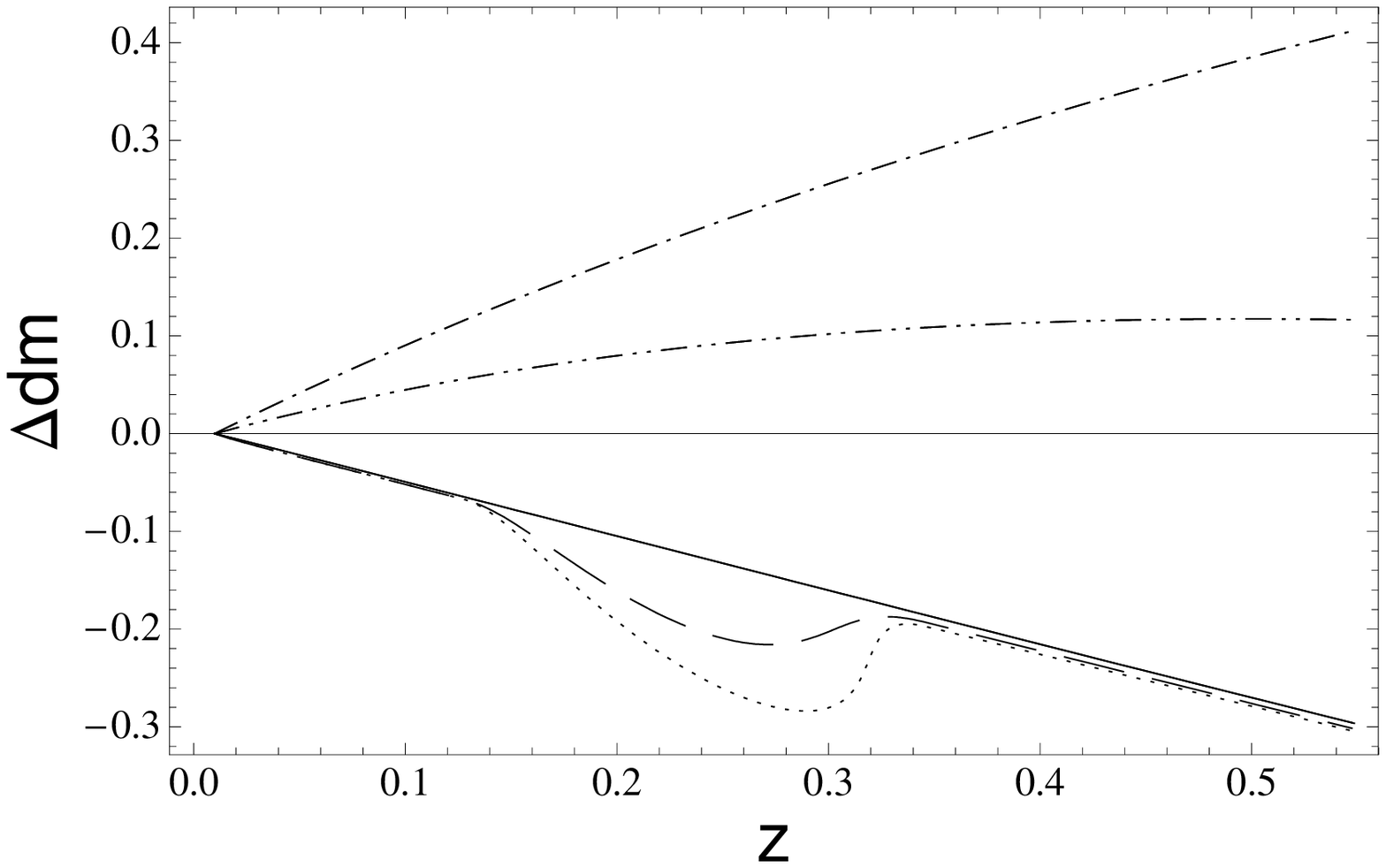,width=8.55cm}
\end{flushright}
\vspace{-10pt}
\caption{The upper panel shows two example density profiles resulting
  from smooth perturbations in $k(r)$ of the form (\ref{kr}), with different depths.  The
  lower panel shows the corresponding distance moduli. The dot-dashed,
  double-dot-dashed and solid lines are as in Figure \ref{fig1}.}
\label{toy2}
\end{figure}

The observer in this space-time is taken to be at the centre of
symmetry, so that the luminosity distance to a source at $(t_e,r_e)$
is given by (\ref{rL}).  In Figure \ref{toy2} we show a couple of
example cases.  The radial profile of the curvature perturbation in
each of these cases is that of (\ref{kr}), with its minimum displaced
from the centre of symmetry.  The upper panel in Figure \ref{toy2}
shows the energy density, as a function of redshift, experienced by the
photon as it travels along our past light cone.  While the
perturbation in $k(r)$ is symmetric, the resulting
energy density distribution can be seen to be highly asymmetric, with
a considerable over-density on the side of the under-dense shell that is
furthest from the observer.

In the lower panel of Figure \ref{toy2} we plot the distance modulus
that would be measured by an observer at the centre of symmetry, for
these two different cases.  It can be seen that the under-density
produces a considerable deviation from the background EdS model.
Comparing this effect with that shown in Figure \ref{fig1}, for the
Swiss Cheese case, one can see that the magnitude of the deviation is
of the same order.  More striking, however, is that the
under-density in the Onion universe causes a very different deviation
in distance modulus.

\label{lastpage}

\end{document}